\documentclass[aps,prd,showpacs,preprintnumbers,nofootinbib]{revtex4}
\usepackage[dvips]{graphicx}
\usepackage{bm,latexsym,amsmath,amssymb,amsfonts}

\begin{document}


\title{Asymptotic structure at timelike infinity: higher orders}

\author{Kentaro Tanabe}
\affiliation{Yukawa Institute for Theoretical Physics, Kyoto University, Kyoto 606-8502, Japan}
\author{Tetsuya Shiromizu}
\affiliation{Department of Physics, Kyoto University, Kyoto 606-8502, Japan}
\begin{abstract}
Bearing the final fate of gravitational collapse in mind, we study the asymptotic 
structures at timelike infinity in four dimensions. 
Assuming that spacetimes are asymptotically stationary, we will examine 
the asymptotic structure of asymptotic stationary spacetimes in a systematic way. 
Then we see that the asymptotic stationarity strongly restricts the asymptotic structure 
at timelike infinity. We also observe that the resulted asymptotic form of the 
metric have the deviation from the Kerr black hole spacetime without assuming of the 
presence of some additional asymptotic symmetries. 
\end{abstract}


\maketitle

\section{introduction}

If gravitational collapse takes place and black holes are formed, we can 
expect that the energies and momenta of fields in the exterior region 
would be radiated away to infinities or fallen into black holes. As a consequence, 
one usually expects that spacetimes would be approaching to asymptotically vacuum, stationary 
state. By virtue of the uniqueness theorem of the Kerr spacetime in four dimensions 
\cite{Israel},
we can regard the Kerr spacetime as this final state and use the Kerr spacetime 
to analyze astrophysical phenomenon around the black hole candidates in the universe. 
However, since this 
uniqueness theorem requires the presence of a timelike Killing vector rigorously, 
we cannot say anything 
about the late time phase of gravitational collapse, 
which is nearly but not completely stationary phase. 
The purpose of this paper is to study how dynamical space-time is approaching to 
the final state (supposed to be the Kerr spacetime) from the view point of asymptotic 
structures. Since we consider spacetimes 
at the late time phases, the $1/t$-expansion would be reliable. 
Hence, we will consider 
the asymptotic structure of space-times at timelike infinity. 

Gen and one of the authors in the present article developed the way to 
investigate the asymptotic structure at timelike infinity 
\cite{Gen:1997az,Gen:1998ay}. Therein the notion of asymptotic flatness at 
spatial infinity formulated by 
Ashtekar and Romano \cite{Ashtekar:1991vb} was used. In Ref. \cite{Gen:1998ay}, the 
first order structure at timelike infinity was studied and the notion of 
asymptotic stationarity was introduced. Roughly speaking, asymptotic stationarity means 
that a spacetime is approximately stationary near timelike infinity, but not 
completely stationary. Then one may expect that asymptotically stationary spacetime at 
timelike infinity describes 
the late time phase of dynamical spacetimes. In Ref. \cite{Gen:1998ay}, it was shown that 
asymptotically stationary, vacuum and flat spacetime is uniquely asymptotically 
Schwarzschild spacetime, i.e., such spacetime always has same first order structure 
as the Schwarzschild spacetime. The first order structure has the information of mass 
of the spacetime. By this theorem, as long as considering the first order in 
$1/t$ expansion, we can say that asymptotically stationary spacetimes 
is approaching to the Schwarzschild spacetime. Then it is natural to expect 
that the spacetimes will be the Kerr black hole spacetime in the next to the leading order.  
In this paper, therefore, we will explore the 
higher order structures at timelike infinity, that is, we will address 
if asymptotically stationary spacetimes always approach to the Kerr 
spacetime. 

The rest of this paper is organized as follows. In Sec. II we will introduce 
the notion of timelike infinity, asymptotic stationarity and first order structure. 
In Sec. III we systematically study 
the $n$-th order structure at timelike infinity focusing on the new degree of freedoms. 
In Sec. IV we discuss the 
second and third order structures in the details. Then we will compare these 
structure with those of the Kerr spacetime in Sec. V. Finally we will 
summarize our work and discuss future issues. 
In Appendix A we introduce the tensor harmonics on three dimensional hyperboloid. 
In Appendix B we give the brief derivation of the zero-th 
and first order structures. In Appendix C we give some useful formulae for the computation 
of the second and third order structures. In Appendix D
we will describe the definition of the 
multipole moments which will be used in Sec. V.

\section{timelike infinity and asymptotic stationarity}
\label{Sec:TI}

In this section, following Ref. \cite{Gen:1997az,Gen:1998ay}, 
we shall introduce the notion of timelike infinity, asymptotic stationarity and first order 
structure. 
In this paper we focus on only vacuum spacetimes. It is easy to extend our work 
to non-vacuum cases if we assume that the matter rapidly decays near 
timelike infinity. 

\subsection{Definition of timelike infinity}

We define timelike infinity in asymptotically flat spacetimes. 
If there is a function $\Omega$ which satisfies following conditions, (i)
%
$\nabla_{\mu}\Omega \neq 0$
%
and (ii) $n^{\mu}:=\Omega^{-4}\nabla^{\mu}\Omega$ and $q_{\mu\nu}:=\Omega^{2}
(g_{\mu\nu}+\Omega^{-4}F^{-1}\nabla_{\mu}\Omega\nabla_{\nu}\Omega)$
admit the smooth limits at $\Omega\,=\,0$ with $q_{\mu\nu}$ having signature $(+++)$, 
we call this spacetime asymptotically flat at timelike infinity. Here 
$\Omega=0$ is the timelike infinity.

Let us suppose that the metric can be expanded near timelike infinity $\Omega=0$ as
%
\begin{eqnarray}
g_{\mu\nu}\,=\,g^{(0)}_{\mu\nu}\,+\sum_{n=1}^{\infty}\,\Omega^{n} g^{(n)}_{\mu\nu}.
\end{eqnarray}
%
By solving the vacuum Einstein equations $R_{\mu\nu}=0$ near timelike infinity, 
we can show \cite{Gen:1997az}
%
\begin{eqnarray}
& & F \hat{=}1,  \\
& & q_{ab}dx^{a}dx^{b} \hat{=}h^{(0)}_{ab}dx^{a}dx^{b}
\equiv d\rho^{2}+\sinh^2{\rho}(d\theta^{2}+\sin^{2}\theta d\phi^{2}),
\end{eqnarray}
%
where $\hat{=}$ denotes the evaluation on timelike infinity $\Omega=0$. See Appendix B for the 
derivation. 
Here we denote $x^a=(\rho,\theta,\phi)$. 
Introducing the coordinate $\eta$ defined by $\eta=\log\Omega$, 
$g^{(0)}_{\mu\nu}dx^{\mu}dx^{\nu}$ becomes 
%
\begin{eqnarray}
g^{(0)}_{\mu\nu}dx^{\mu}dx^{\nu}\,=\,e^{-2\eta}\left[
-d\eta^{2}+d\rho^{2}+\sinh^2{\rho}(d\theta^{2}+\sin^{2}{\theta} d\phi^{2})
\right]. \label{zero}
\end{eqnarray}
%
$g^{(0)}_{\mu\nu}$ gives us the metric of 
the zero-th order structure at timelike infinity. In this coordinate timelike infinity is located 
at $\eta=-\infty$. 
By coordinate transformation $t=\Omega^{-1}\cosh{\rho}$ and 
$r=\Omega^{-1}\sinh{\rho}$, it is shown that the zero-th 
order structure metric is the well-known Minkowski one, that is, 
%
\begin{eqnarray}
g^{(0)}_{\mu\nu}dx^{\mu}dx^{\nu}=-dt^2+dr^2+r^2(d\theta^{2}+\sin^{2}{\theta} d\phi^{2}).
\end{eqnarray}
%

\subsection{Asymptotic stationarity}

Obviously the stationary Killing vector at the zero-th order is given by 
%
\begin{gather}
\xi^{(0)}=-\frac{\partial}{\partial t}=\Omega\cosh{\rho}\frac{\partial}{\partial\eta}
 +\Omega\sinh{\rho}\frac{\partial}{\partial\rho}.
\end{gather}
%
Its dual vector is 
%
\begin{gather}
\xi^{(0)*}=-\Omega^{-1}\cosh{\rho} d\eta
 +\Omega^{-1}\sinh{\rho} d \rho.
\end{gather}
%

Then one may introduce the asymptotic stationarity as follows. If spacetime 
$(\mathcal{M},g_{\mu\nu})$ has a vector field $\xi$ which satisfies the 
following conditions
%
\begin{gather}
\xi^{\mu}\xi_{\mu}\hat{=}-1, \label{xicond1}\\
\left(\mbox \pounds_{\xi}g \right) _{\mu\nu}\,=\,O(\Omega^{n}) 
\label{xicond2}, 
\end{gather}
%
the space-time is asymptotic stationary at order $n$ and we call the vector field $\xi$ 
asymptotic Killing vector at order $n$. Here we note that the order of $\Omega$ is 
different from that in Ref. \cite{Gen:1998ay}. This is due to different basis from that 
in Ref. \cite{Gen:1998ay}. 
In Ref. \cite{Gen:1998ay}, the conventional tetrad was used. On the other hand, we use 
the coordinate system of ($\eta,\rho,\theta,\phi$). 

\subsection{First order structure}

Now we can discuss the first order structure. See Appendix B for the details. 
In general $g^{(1)}_{\mu\nu}$ can be expressed as
%
\begin{eqnarray}
g^{(1)}_{\mu\nu}\,=\,e^{-2\eta}\left[
-F^{(1)}d\eta^{2}+2\beta^{(1)}_{a}d\eta dx^{a}+h^{(1)}_{ab}dx^{a}dx^{b}
\right].
\end{eqnarray}
%
Solving the vacuum Einstein equations $R_{\mu\nu}=0$ and imposing 
the asymptotic stationary condition at the first order, we can obtain 
the solutions up to the first order as \cite{Gen:1998ay}
%
\begin{eqnarray}
(g^{(0)}_{\mu\nu}+\Omega g^{(1)}_{\mu\nu})dx^{\mu}dx^{\nu}\,&=&\,e^{-2\eta}
\left[
\Bigl(-1+2m\frac{\cosh{2\rho}}{\sinh{\rho}}\Omega \Bigr) d\eta^{2}-8m\cosh{\rho}\Omega d\eta d\rho
\right. 
\notag \\
&&~~~~~~~~~~~~~~~~~~~~~~~~~~~~~~~~~
+\Bigl(1+2m\frac{\cosh{2\rho}}{\sinh{\rho}}\Omega \Bigr) d\rho^{2} 
+\sinh^{2}{\rho}(d\theta^{2}+\sin^{2}{\theta}d\phi^{2})
\bigg],\label{aSch}
\end{eqnarray}
%
where $m$ is a constant which would be proportional to the black hole mass. 
We emphasize that $F^{(1)}$ is gauge invariant. Due to the presence of the asymptotic stationarity 
it is shown that it has only $l=0$ mode and then we had the result of Eq. (\ref{aSch}). 
See Appendix B for the details. 

At the first order the asymptotic Killing vector is given by 
%
\begin{gather}
\xi_{\mu}\,=\,\Omega^{-1}\xi^{(0)}_{\mu} + \xi_{\mu}^{(1)},
\end{gather}
%
where
%
\begin{gather}
\xi^{(1)}\,=\,2m\coth{\rho}d\eta -2m d\rho.
\end{gather}
%
This first order structure has the same as that of the Schwarzschild 
spacetime with the mass $m$. Thus, asymptotically stationary spacetimes near the timelike infinity 
can be approximated by the Schwarzschild spacetime up to the first order.

\section{$n$-th order structure at timelike infinity}
\label{Sec:nO}

In this section we further consider the $n$-th order structure at timelike infinity. 
Near timelike infinity, we expand the metric up to the $n$-th order as
%
\begin{eqnarray}
g_{\mu\nu}\,=\,\sum_{i=0}^{n}\Omega^{i}g^{(i)}_{\mu\nu} +O(\Omega^{n+1}),
\end{eqnarray}
%
where $g^{(i)}_{\mu\nu}$ is the metric at the $i$-th order. In previous section 
the zero-th and first structures were briefly reviewed. From now on,  
we will examine the higher order structure recursively. 

\subsection{The Einstein equation}

First we will consider the Einstein equation for the metric of the $n$-th order structure 
$g^{(n)}_{\mu\nu}$. In general we can write down the $n$-th order parts of 
the metric as
%
\begin{eqnarray}
g^{(n)}_{\mu\nu}dx^{\mu}dx^{\nu}\,=\,
e^{-2\eta}[-F^{(n)}(x^{a})d\eta^{2}-2\beta^{(n)}_{a}(x^{b})d\eta dx^{a}
+h^{(n)}_{ab}(x^{a})dx^{a}dx^{b} ].
\end{eqnarray}
%
$h^{(n)}_{ab}$ can be decomposed into the trace and traceless parts as 
%
\begin{eqnarray}
h^{(n)}_{ab}\,=\,\psi^{(n)}h^{(0)}_{ab}+\chi^{(n)}_{ab}.
\end{eqnarray}
%
Then, the Einstein equation for $g^{(n)}_{\mu\nu}$ becomes
%
\begin{gather}
(D^{2}-3n)F^{(n)}+3n(1-n)\psi^{(n)}+2(n-1)D^{a}\beta^{(n)}_{a}
\,=\,S^{(n)}[g^{(1)},\cdots ,g^{(n-1)}], \\
(D^{2}-2)\beta^{(n)}_{a}+2D_{a}F^{(n)}+2nD_{a}\psi^{(n)}-nD^{b}\chi_{ab}^{(n)}
\,=\,S^{(n)}_{a}[g^{(1)},\cdots ,g^{(n-1)}], \\
(D^{2}+3-(n-1)^2)\chi_{ab}^{(n)}+D_{a}D_{b}F^{(n)}+(4-n)F^{(n)}h^{(0)}_{ab}
+D_{a}D_{b}\psi^{(n)}-(n-4)(n-1)\psi^{(n)}h^{(0)}_{ab}
\notag \\
~~~~~~~~~~~~~~~~~~~~~~~~~~~~~~~~~~~~~
+D^{2}\psi^{(n)}h^{(0)}_{ab}-2D^{m}\beta^{(n)}_{m}h_{ab}^{(0)}
+2(n-2)D_{(a}\beta_{b)}^{(n)}
\,=\,S^{(n)}_{ab}[g^{(1)},\cdots ,g^{(n-1)}].
\end{gather}
%
$S^{(n)}$, $S^{(n)}_{a}$ and $S^{(n)}_{ab}$ are source terms written by 
the lower order quantities $g^{(1)},\cdots,g^{(n-1)}$ in non-linear form. 
We will not write down 
the expressions explicitly because they are not important in the analysis of this section. 
But, we will do so for the second and third order structures in the next section. 
$D_{a}$ is the covariant derivative with respect to $h^{(0)}_{ab}$ and $D^{2}=D^{a}D_{a}$. 

Since the equations are linear with respect to $g^{(n)}_{\mu \nu}$, we decompose 
$g^{(n)}_{\mu\nu}$ into the homogeneous parts $g^{(n),\text{hom}}_{\mu\nu}$ 
and the source parts $g^{(n),\text{sou}}_{\mu\nu}$. 
The source part consists of only lower order quantities. On the other hand,  
the homogeneous parts have new degree of freedoms in which we are interested 
in the $n$-th order. In the following argument, therefore, we will focus on 
only the homogeneous terms. For the brevity we will skip the index ``hom" for the 
homogeneous parts in the remaining part of this subsection. 

The metric of the $n$-th order structures, $g^{(n)}_{\mu\nu}$,  has the gauge freedom generated by 
$x^{\mu}\rightarrow x^{\mu}+k^{\mu}$ with $k_{\mu}dx^{\mu}=\Omega^{n-2}(Td\eta+L_{a}dx^{a})$. 
Of course, the lower order structures 
also have the gauge freedoms which affects on the $n$-th order structures. 
However, these gauge transformations at lower orders 
act only on the source term $g^{(n),\text{sou}}_{\mu\nu}$. 
Hence, we consider only the gauge freedoms at the pure $n$-th order. 
The metric is transformed by 
this gauge transformation as
%
\begin{gather}
F^{(n)}\rightarrow F^{(n)}-2(n-1)T\,,\,
\beta_{a}^{(n)}\rightarrow \beta_{a}^{(n)}+D_{a}T+nL_{a} \label{ngauge1}\\
\psi^{(n)}\rightarrow \psi^{(n)}+2T+2D^{a}L_{a}\,,\,
\chi^{(n)}_{ab}\rightarrow \chi^{(n)}_{ab}+2D_{(a}L_{b)}-\frac{2}{3}h^{(0)}_{ab}D^{m}L_{m}.
\label{ngauge2}
\end{gather}
%
Then we can take the ``Poisson gauge" as 
%
\begin{eqnarray}
&  & D^{a}\beta_{a}^{(n)}=0, \\
&  &D^{b}\chi^{(n)}_{ab}=0.
\end{eqnarray}
%
Note that we imposed the Poisson gauge on only homogeneous terms. There are still gauge 
freedoms generated by $T$ and $L_{a}$ satisfying
%
\begin{eqnarray}
& & D^{2} T +nD^{a}L_{a}=0, \\
& & (D^{2} -2)L_{a}-\frac{1}{3}D_{a}D^{m}L_{m} =0.
\end{eqnarray}
%
Then the Einstein equations for the homogeneous terms become
%
\begin{gather}
(D^{2}-3n)F^{(n)}+3n(1-n)\psi^{(n)}\,=\,0, \label{ein1}\\
(D^{2}-2)\beta_{a}^{(n)}+2D_{a}(F^{(n)}+n\psi^{(n)})\,=\,0, \label{ein2} \\
(D^{2}+3-(n-1)^2)\chi^{(n)}_{ab}+(D_{a}D_{b}+(n-4)h^{(0)}_{ab})F^{(n)}
+(D_{a}D_{b}-(n-4)(n-1)h^{(0)}_{ab}+D^{2}h^{(0)}_{ab})\psi^{(n)}
+2D_{(a}\beta^{(n)}_{b)}\,=\,0. \label{ein3}
\end{gather}
%
Adding the trace part of Eq. (\ref{ein3}) to Eq. (\ref{ein1}), we can obtain 
%
\begin{eqnarray}
D^{2}(F^{(n)}+n\psi^{(n)})\,=\,0.
\end{eqnarray}
%
Taking $T=F^{(n)}+n\psi^{(n)}$ and $L_{a}=0$, which keep the 
Poisson gauge, we can always set 
%
\begin{eqnarray}
F^{(n)}+n\psi^{(n)}=0. 
\end{eqnarray}
%
There are still the residual gauge 
generated by $L_{a}$ satisfying $(D^{2}-2)L_{a}=0$. 
Since the equation for $\beta^{(n)}_{a}$ is
%
\begin{eqnarray}
(D^{2}-2)\beta_{a}^{(n)}\,=\,0,
\end{eqnarray}
%
we can set $\beta_{a}^{(n)}=0$ by this residual gauge. 
Then the remaining equations we should solve are
%
\begin{gather}
(D^{2} -3)F^{(n)}\,=\,0, \\
(D^{2} +3-(n-1)^2)\chi_{ab}^{(n)}\,=\,-\frac{(n-1)}{n}D_{a}D_{b}F^{(n)}
+\frac{n-1}{3n}D^{2} F^{(n)}h^{(0)}_{ab}.
\end{gather}
%
Let us consider the gauge transformation generated by $T=F^{(n)}/(2(n-1))$ and 
$nL_{a}=-D_{a}T$ which also keeps the Poisson gauge. Then, in this gauge, we can see  
%
\begin{eqnarray}
\beta_{a}^{(n)}=F^{(n)}=0
\end{eqnarray}
%
holds.  
Note that the $n=1$ case is exceptional because $F^{(1)}$ is already gauge invariant. 
Thus we cannot eliminate $F^{(1)}$ in the $n=1$ case. As stated in the previous section, 
the degree of freedom of the first order structure is included in 
$F^{(1)}$ (See Appendix B for the details.). 

At the $n$-th orders($n>1$), $F^{(n)}$ is pure gauge modes. The degree of freedom at the 
$n$-th order are contained in $\chi^{(n)}_{ab}$ satisfying 
%
\begin{eqnarray}
(D^{2}+3-(n-1)^2 )\chi_{ab}^{(n)}\,=\,0.
\end{eqnarray}
%
The solutions to the above can be written by the tensor harmonic on the hyperboloid, 
$\mathcal{G}^{+,plm}_{ab}$ and $\mathcal{G}_{ab}^{-,plm}$(see Appendix A), as 
%
\begin{eqnarray}
\chi_{ab}^{(n)}\,=\,\sum_{l,m}a_{lm}^{(n)}\mathcal{G}_{ab}^{+,n-1\,lm}
+b_{lm}^{(n)}\mathcal{G}_{ab}^{-,n-1\,lm},
\end{eqnarray}
%
where $(+)$-modes have the even parity and $(-)$-modes have the odd parity.
$a_{lm}^{(n)}$ and $b_{lm}^{(n)}$ are constant parameters which 
appear as the new degree of freedoms at the $n$-th order structure. 
In the next subsection, we will look at 
how asymptotic stationarity restricts the above new degree.

\subsection{Asymptotic stationarity}

Next, we consider the condition of asymptotic stationarity.
We impose this condition on the $n$-th order structure. Asymptotically 
stationary condition for the $n$-th order structures is 
%
\begin{eqnarray}
\nabla_{\mu}\xi_{\nu}+\nabla_{\nu}\xi_{\mu}\,=\,O(\Omega^{n}) \label{nAS},
\end{eqnarray}
%
where $\xi$ is supposed to be an asymptotic stationary Killing vector and expanded as follows
%
\begin{eqnarray}
\xi_{\mu}\,=\,\sum_{i=0}^{n}\Omega^{i-1}\xi_{\mu}^{(i)} +O(\Omega^{n}).
\end{eqnarray}
%
In general we can decompose $\xi_\mu^{(n)}$ into two parts, $\xi^{(n),{\rm hom}}_\mu$ and $\xi^{(n),{\rm sou}}$ 
which are associated with the homogeneous parts only of $n$-th order and others respectively. 

From the $(\eta,\eta)$-components of Eq. (\ref{nAS}) we first see
%
\begin{eqnarray}
\partial_\eta (\xi_\eta^{(n)}\Omega^{n-1})-{}^{(0)}\Gamma^{\mu}_{\eta \eta}\xi_\mu^{(n)} \Omega^{n-1}-
{}^{(1)}\Gamma^\mu_{\eta\eta}\xi_\mu^{(n-1)}\Omega^{n-2}-\cdots - {}^{(n)}\Gamma^\mu_{\eta\eta}\xi^{(0)}_\mu \Omega^{-1}
=O(\Omega^{n}),
\end{eqnarray}
%
where $\Gamma^\alpha_{\mu\nu}
=(1/2)g^{\alpha\beta}(\partial_\mu g_{\beta\nu}+\partial_\nu g_{\beta\nu}-\partial_\beta g_{\mu\nu})
={}^{(0)}\Gamma^\alpha_{\mu\nu}+{}^{(1)}\Gamma^\alpha_{\mu\nu}
+ {}^{(2)}\Gamma^\alpha_{\mu\nu}+\cdots$ and the $\Omega$-dependences  are included in 
each terms. If one focuses on $\xi^{(n),{\rm hom}}_\eta$, the above equation is simplified to 
%
\begin{eqnarray}
\partial_\eta (\xi_\eta^{(n),{\rm hom}}\Omega^{n-1})-{}^{(0)}\Gamma^{\mu}_{\eta \eta}\xi_\mu^{(n),{\rm hom}} 
\Omega^{n-1}=O(\Omega^{n}).
\end{eqnarray}
%
After short computations, it gives us  
%
\begin{eqnarray}
n \xi_{\eta}^{(n),{\rm hom}}=0.
\end{eqnarray}
%

Using this the ($\eta,a$)-components of Eq. (\ref{nAS}) for $\xi_a^{(n),{\rm hom}}$ becomes 
%
\begin{eqnarray}
\partial_\eta (\xi_a^{(n),{\rm hom}}\Omega^{n-1})-2{}^{(n)}\Gamma_{a \eta}^{\mu, {\rm hom}} 
\xi_\mu^{(0)} \Omega^{-1}
-2{}^{(0)}\Gamma_{a\eta}^{\mu} \xi_{\mu}^{(n), {\rm hom}}\Omega^{n-1}=O(\Omega^n).
\end{eqnarray}
%
After short calculation, we see 
%
\begin{eqnarray}
(n+1)\xi_a^{(n), {\rm hom}}-n\chi^{b(n), {\rm hom}}_a \xi_b^{(0), {\rm hom}}=0,
\end{eqnarray}
%
and then 
%
\begin{eqnarray}
\xi_{a}^{(n),{\rm hom}}=\frac{n}{n+1}\chi^{b(n), {\rm hom}}_{a}\xi_b^{(0),{\rm hom}}. 
\end{eqnarray}
%

Then the ($a,b$)-components imply us the condition on $\chi_{ab}^{(n), {\rm hom}}$ as 
%
\begin{eqnarray}
2D_{(a}\chi_{b)\rho}^{(n), {\rm hom}}-(n+1)D_{\rho}\chi_{ab}^{(n), {\rm hom}}-(n+2)(n-1)\coth{\rho}
\chi^{(n), {\rm hom}}_{ab}\,=\,0 
\label{nab}.
\end{eqnarray}
%
This is the asymptotically stationary condition on the $n$-th order structure. We will 
look at this precisely. The $(\rho,\rho)$ and $(\rho,A)$-components of the above become
%
\begin{gather}
a_{l,m}^{(n)}\left(
\frac{\partial}{\partial \rho}P^{n-1,l}+n\coth{\rho}P^{n-1,l}
\right)Y^{l,m}(\theta,\phi) =0,\\
b_{l,m}^{(n)}\left(
\frac{\partial}{\partial \rho}P^{n-1,l}+n\coth{\rho}P^{n-1,l}
\right)\mathcal{D}_{A}Y^{l,m}(\theta,\phi)=0,
\end{gather}
%
where $\mathcal{D}_{A}$ is the derivative with $x^{A}=(\theta,\phi)$
and $Y^{l,m}$ is the spherical harmonics on $S^{2}$. 
It is easy to see that the non-trivial solution is only permitted for the cases of $l=n-1$. 
Then the solution is proportional to $P^{n-1,n-1}=1/\sinh^n{\rho}$. It can be easily confirmed 
that $l=n-1$ modes satisfies the $(A,B)$-components of Eq. (\ref{nab}) too. 

Therefore, the degree of freedom of stationary $n$-th order 
structure is only $l=n-1$ modes. Remember that the asymptotic stationarity could induce the 
axisymmetricity at the first order. However, it could not guarantee the asymptotic 
axisymmetry at higher orders, that is, $m \neq 0$ modes exist. 
As seen later in Sec. V, $m\neq 0$ modes imply the deviations from the Kerr black hole 
spacetime in the higher multipole moments. 

Here we focused on the homogeneous parts and then we could observe that the asymptotic 
Killing equation at $n$-th order 
affects the homogeneous parts of the $n$-th order structure of timelike infinity. 
However, one wonders if the ($a,b$)-component of Eq. (\ref{nAS}) gives us the additional constraints on 
the homogeneous parts of the lower order structures than $n-1$. This issue will be occurred once one thinks of 
``source" parts of the asymptotic Killing equation.  Although any additional constrains does not appear intuitively, 
this is non-trivial issue. In the next section, we will confirm that this property holds 
in concrete examples. Then we would expect that this is 
valid for all $n$-th orders. 

\section{Second and Third order structures}

In this section we will look at the second and third order structures in the details. 
We will also consider the contributions from the source terms explicitly which were not addressed in 
the previous section.

\subsection{Second order structure}

At the second order, the Einstein equations are
%
\begin{gather}
(D^2 -6)F^{(2)} - 6\psi^{(2)} + 2D^{a}\beta^{(2)}_{a}
\,=\,-\frac{16m^{2}}{\sinh^2{\rho}}, \label{2ndEin1}\\
(D^2 -2)\beta^{(2)}_{a} - 2D^{b}\chi_{ab}^{(2)}+2D_{a}F^{(2)}
+ 4D_{a}\psi^{(2)}\,=\,-16m^2\frac{\coth{\rho}}{\sinh^{2}{\rho}}h^{(0)}_{a\rho}, \label{2ndEin2}\\
(D^2 +2)\chi^{(2)}_{ab}-2D_{(a}D^{c}\chi_{b)c}^{(2)}+(D_{a}D_{b}+2h^{(0)}_{ab})F^{(2)}
+(D_{a}D_{b}+h_{ab}^{(0)}D^2 +2h^{(0)}_{ab})\psi^{(2)}
-2h^{(0)}_{ab}D^{m}\beta^{(2)}_{m}
\notag \\
~~~~~~~~~~~~~~~~~~~~~~~~~~~~~~~~~~~~~~~~~~~~~~~~~~~~~~~~~~~~~~~~~~~
\,=\,16m^{2}\frac{\coth^{2}{\rho}}{\sinh^{2}{\rho}}h_{a\rho}^{(0)}h_{b\rho}^{(0)} \label{2ndEin3}.
\end{gather}
%
The right-hand sides come from the source parts composed of the lower order quantities. 
Now we decompose the aboves into the homogeneous and source parts. The non-trivial 
solutions of the source parts become
%
\begin{gather}
F^{(2),\text{sou}}\,=\,-4m^{2}, \\
\beta^{(2),\text{sou}}_{\rho}\,=\,-4m^{2}\coth{\rho}, \\
h^{(2),\text{sou}}_{\rho\rho}\,=\,4m^{2}\coth^{2}{\rho},
\end{gather}
%
and the other components of $g^{(2),\text{sou}}_{\mu\nu}$ vanish. 
The homogeneous solutions is
%
\begin{gather}
\chi_{ab}^{(2),\text{hom}}\,=\,\sum_{lm}a^{(2)}_{lm}\mathcal{G}^{+, 1lm}_{ab}
+b^{(2)}_{lm}\mathcal{G}^{-, 1lm}_{ab},
\end{gather}
%
where the summation is taken over $l\in{\bf{Z}}$, $l\geq 1$ and $|m|\leq l$. 
The other homogeneous terms can be vanished by the appropriate gauge choices as shown in 
Sec. III. 

Next we consider the asymptotic stationarity at the second order. 
Spacetimes are asymptotically stationary at the second order when there is the 
asymptotic Killing vector satisfying 
%
\begin{eqnarray}
\nabla_{\mu}\xi_{\nu}+\nabla_{\nu}\xi_{\mu}\,=\,O(\Omega^{2}). \label{2AS}
\end{eqnarray}
%
$\xi$ can be expanded near timelike infinity as
%
\begin{eqnarray}
\xi_{\mu}\,=\,\Omega^{-1}\xi^{(0)}_{\mu}+\xi^{(1)}_{\mu}+\Omega \xi^{(2)}_{\mu}
+O(\Omega^{2}).
\end{eqnarray}
%
Since Eq. (\ref{2AS}) is linear with respect to $\xi^{(2)}$, we can decompose 
$\xi^{(2)}$ into the homogeneous and source parts. 

The homogeneous part, $\xi^{(2),\text{hom}}$, can be written as 
%
\begin{eqnarray}
& & \xi^{(2),\text{hom}}_{\eta}=0, \\
& & \xi^{(2),\text{hom}}_{a}=\frac{2}{3}\chi^{(2),\text{hom}}_{ab}\xi^{(0)b}.
\end{eqnarray}
%
As seen in the previous section, 
the ($a,b$)-components of Eq. (\ref{2AS}) show that 
$\chi^{(2),\text{hom}}_{ab}$ can have only $l=1$ modes and its explicit 
form of $\chi^{(2),\text{hom}}_{ab}$ is given by 
%
\begin{eqnarray}
\chi_{ab}^{(2),\text{hom}}\,=\,
\begin{pmatrix}
\frac{A^{(2)}}{\sinh^{4}{\rho}} &
-\frac{1}{2}\frac{\coth{\rho}}{\sinh^{2}{\rho}}\mathcal{D}_{A}A^{(2)}
-\frac{1}{\sinh^{2}{\rho}}\epsilon_{A}^{~B}\mathcal{D}_{B}B^{(2)} \\
-\frac{1}{2}\frac{\coth{\rho}}{\sinh^{2}{\rho}}\mathcal{D}_{A}A^{(2)}
-\frac{1}{\sinh^{2}{\rho}}\epsilon_{A}^{~B}\mathcal{D}_{B}B^{(2)} &
-\frac{1}{2}A^{(2)}\sigma_{AB} 
\end{pmatrix},
\end{eqnarray}
%
where 
%
\begin{eqnarray}
&  &A^{(2)}=a_{x}\sin\theta\cos\phi +a_{y}\sin\theta\sin\phi +a_{z}\cos\theta,\\
& & B^{(2)}=b_{x}\sin\theta\cos\phi +b_{y}\sin\theta\sin\phi +b_{z}\cos\theta.
\end{eqnarray}
%
$\sigma_{AB}$ and $\epsilon_{AB}$ are the metric of the unit sphere and 
the antisymmetric tensor 
with respect to $\sigma_{AB}$. 

Let us consider the source term $\xi^{(2),\text{sou}}$ which 
can be determined by the $(\eta,\mu)$-component in Eq. (\ref{2AS})
%
\begin{eqnarray}
\partial_\mu (\xi_\nu^{(2), {\rm sou}}\Omega)+
\partial_\nu (\xi_\mu^{(2), {\rm sou}}\Omega)
-2{}^{(2)}\Gamma^{\alpha, {\rm sou}}_{\mu\nu}\xi_\alpha^{(0)}\Omega^{-1}
-2\Gamma^{(1)}_{\mu\nu}\xi_\alpha^{(1)}-2{}^{(0)}\Gamma^\alpha_{\mu\nu}\xi^{(2), {\rm sou}}_\alpha \Omega=O(\Omega^2),
\label{ak2}
\end{eqnarray}
%
where ${}^{(2)}\Gamma^{\alpha, {\rm sou}}_{\mu\nu}$ is the part which contains only 
the source parts. After short calculation we can see 
%
\begin{eqnarray}
{}^{(2)}\Gamma^{\alpha, {\rm sou}}_{\eta \nu}\xi_\alpha^{(0)}\Omega^{-1}
+\Gamma^{(1)}_{\eta \nu}\xi_\alpha^{(1)}=0. 
\end{eqnarray}
%
See some useful formulae in Appendix C for concrete calculations. 
Then Eq. (\ref{ak2}) implies 
%
\begin{eqnarray}
\xi^{(2), {\rm sou}}_\mu=0.
\end{eqnarray}
%
Then we can see that 
%
\begin{eqnarray}
{}^{(2)}\Gamma^{\alpha, {\rm sou}}_{ab}\xi_\alpha^{(0)}\Omega^{-1}
+\Gamma^{(1)}_{ab}\xi_\alpha^{(1)}=0. 
\end{eqnarray}
%
holds and then the ($a,b$)-component of Eq. (\ref{ak2}) becomes to be trivial. 
Thus there are no additional constraints on the metric. 

To be summarized, the second order structure which is asymptotically stationary at 
the second order is 
%
\begin{eqnarray}
& & F^{(2)}=-4m^{2},\\
& & \beta_{\rho}=-4m^{2}\coth{\rho}, \\
& & h^{(2)}_{ab}=
\begin{pmatrix}
4m^{2}\coth^{2}{\rho}+\frac{A^{(2)}}{\sinh^{4}{\rho}} &
-\frac{1}{2}\frac{\coth{\rho}}{\sinh^{2}{\rho}}\mathcal{D}_{A}A^{(2)}
-\frac{1}{\sinh^{2}{\rho}}\epsilon_{A}^{~B}\mathcal{D}_{B}B^{(2)} \\
-\frac{1}{2}\frac{\coth{\rho}}{\sinh^{2}{\rho}}\mathcal{D}_{A}A^{(2)}
-\frac{1}{\sinh^{2}{\rho}}\epsilon_{A}^{~B}\mathcal{D}_{B}B^{(2)} &
-\frac{1}{2}A^{(2)}\sigma_{AB} 
\end{pmatrix},
\end{eqnarray}
%
and it has the asymptotic Killing vector
%
\begin{eqnarray}
& & \xi_{\eta}=\Omega^{-1}\cosh{\rho}+2m\coth{\rho}+O(\Omega^{2}), \\
& & \xi_{a}=\Omega^{-1}\xi_{a}^{(0)}+\xi^{(1)}_{a}
+\Omega\frac{2}{3}\chi^{(2),\text{hom}}_{ab}\xi^{(0)b}
+O(\Omega^{2}).
\end{eqnarray}
%

\subsection{Third order structure}

Next we shall consider the third order structures in the details. 
For the convenience it is better to perform the gauge transformation 
$x^{\mu}\rightarrow 
x^{\mu}+\Omega k^{(2)\mu}$ as
%
\begin{eqnarray}
& & k^{(2)}_{\eta} = \frac{1}{6}\frac{A^{(2)}}{\sinh^{2}{\rho}}, \\
& & k^{(2)}_{\rho} = \frac{1}{6}\frac{\coth{\rho}}{\sinh^{2}{\rho}}A^{(2)}, \\
& & k^{(2)}_{A} = -\frac{1}{12}\coth^{2}{\rho}\mathcal{D}_{A}A^{(2)}
-\frac{1}{3}\epsilon_{A}^{~B}\mathcal{D}_{B}B^{(2)}.
\end{eqnarray}
%
They make the computation simple  at the third order. Indeed, as seen later, 
the asymptotic Killing vector is written as $\xi =-\partial_t$ up to the second order 
in this gauge. 
After this gauge transformation the metric at the second order are changed to 
%
\begin{gather}
F^{(2)}\,=\,-4m^{2}-\frac{1}{3}\frac{A^{(2)}}{\sinh^{2}{\rho}} \,,\,
\beta^{(2)}_{\rho}\,=\,-4m^{2}\coth{\rho} \,,\,
\beta^{(2)}_{A}\,=\,-\frac{1}{6}\mathcal{D}_{A}A^{(2)}-\frac{2}{3}\epsilon_{A}^{~B}\mathcal{D}_{B}B^{(2)} 
\label{mod2nd1} \\
h^{(2)}_{\rho\rho}\,=\,4m^{2}\coth^{2}{\rho}-\frac{1}{3}\frac{A^{(2)}}{\sinh^{2}{\rho}} \,,\,
h^{(2)}_{A\rho}\,=\,\frac{1}{6}\coth{\rho}\mathcal{D}_{A}A^{(2)}+\frac{2}{3}\epsilon_{A}^{~B}
\mathcal{D}_{B}B^{(2)} \,,\,
h^{(2)}_{AB}\,=\,\frac{5}{6}A^{(2)}\sigma_{AB}.
\label{mod2nd2}
\end{gather}
%
The Einstein equations at the third order in this gauge are 
%
\begin{gather}
(D^2 -9)F^{(3)}-18\psi^{(3)}+4D^{a}\beta^{(3)}_{a}\,=\,-\frac{8(1-\sinh^{2}{\rho})}{3}mA^{(2)}, \\
(D^2 -2)\beta^{(3)}_{a} -2D^{b}\chi_{ab}^{(3)}+2D_{a}F^{(3)}+6D_{a}\psi^{(3)}
\,=\,S^{(3)}_{a}, \\
(D^2 -1)\chi^{(3)}_{ab}-2D_{(a}D^{c}\chi_{b)c}^{(3)}+2D_{(a}\beta^{(3)}_{b)}-2h^{(0)}_{ab}
D^{m}\beta_{m}^{(3)}+(D_{a}D_{b}+h^{(0)}_{ab})F^{(3)}+(h^{(0)}_{ab}D^2 +2h^{(0)}_{ab}
+D_{a}D_{b})\psi^{(3)}\,=\,S^{(3)}_{ab},
\end{gather}
%
where $S_{a}^{(3)}$ and $S_{ab}^{(3)}$ are 
%
\begin{eqnarray}
& & S_{\rho}^{(3)}=-\frac{8}{3}\frac{\coth\rho}{\sinh^{3}\rho}mA^{(2)},\\
& & S_{A}^{(3)}\,=\,\frac{2m}{\sinh^{3}\rho}\mathcal{D}_{A}A^{(2)}, \\
& & S_{\rho\rho}^{(3)}=\frac{8}{3}\frac{2+\sinh^{2}\rho}{\sinh^{5}\rho}mA^{(2)}, \\
& & S_{\rho A}^{(3)}=-2m\frac{\coth{\rho}}{\sinh^{3}\rho}\mathcal{D}_{A}A^{(2)}, \\
& & S_{AB}^{(3)}=-\frac{8}{3}\frac{mA^{(2)}}{\sinh^{3}\rho}\sigma_{AB}.
\end{eqnarray}
%
As in the second order, we decompose the metric to the homogeneous 
and source parts. The solutions to the source parts are given by 
%
\begin{eqnarray}
& & F^{(3),\text{sou}}=\frac{2}{3}\frac{mA^{(2)}}{\sinh^{3}\rho}-\frac{8m^{3}}{\sinh\rho},\\
& & \beta_{\rho}^{(3),\text{sou}}=-8m^{3}\frac{\coth{\rho}}{\sinh\rho},\\
& & h^{(3),\text{sou}}_{\rho\rho}=\frac{2}{3}\frac{mA^{(2)}}{\sinh^{3}\rho}
+8m^{3}\frac{\coth^{2}\rho}{\sinh^{3}\rho},\\
& & h_{AB}^{(3),\text{sou}}=-\frac{4}{9}\frac{mA^{(2)}}{\sinh{\rho}}\sigma_{AB},
\end{eqnarray}
%
and the homogeneous terms are
%
\begin{gather}
\chi^{(3),\text{hom}}_{ab}\,=\,\sum_{l,m}a^{(3)}_{lm}\mathcal{G}^{+, 2lm}
+b^{(3)}_{lm}\mathcal{G}^{-, 2lm},
\end{gather}
%
where the summation is taken over $l\in \bf{Z}$, $l\geq 2$ and $|m|\leq l$. 
Taking the appropriate gauge as shown in Sec. III, we can eliminate 
the homogeneous parts of the other components of the metric at the third order.

Next we impose the asymptotic stationarity on the third order structure. 
The asymptotic Killing vector we supposed satisfies 
%
\begin{eqnarray}
\nabla_{\mu}\xi_{\nu}+\nabla_{\nu}\xi_{\mu}\,=\,O(\Omega^{3}). \label{3AS}
\end{eqnarray}
%
$\xi$ can be expanded near timelike infinity as
%
\begin{eqnarray}
\xi_{\mu}\,=\,\Omega^{-1}\xi^{(0)}_{\mu}+\xi^{(1)}_{\mu}+\Omega \xi^{(2)}_{\mu}
+\Omega^{2} \xi^{(3)}_{\mu}+O(\Omega^{3}).
\end{eqnarray}
%
Since we performed gauge transformation for the second order structure, 
$\xi^{(2)}$ is also changed to
%
\begin{eqnarray}
& & \xi^{(2)}_{\eta}=\frac{1}{3}\frac{\coth{\rho}}{\sinh{\rho}}A^{(2)}, \\
& & \xi^{(2)}_{\rho}=-\frac{1}{3}\frac{A^{(2)}}{\sinh{\rho}}, \\
& & \xi^{(2)}_{A}=-\frac{2}{3}\frac{\epsilon_{A}^{~B}\mathcal{D_{B}}B^{(2)}}{\sinh{\rho}}.
\end{eqnarray}
%
It is easy to check that the asymptotic Killing vector is simply written by 
$\xi=-\partial_t$ up to the second orders. 
As in the second order structure, we decompose $\xi^{(3)}$ into the homogeneous and source parts. 
For the homogeneous part the $(\eta,\mu)$-components in Eq. (\ref{3AS}) implies 
%
\begin{eqnarray}
& & \xi^{(3),\text{hom}}_{\eta}=0,\\
& & \xi^{(3),\text{hom}}_{a}=\frac{3}{4}\chi^{(3),\text{hom}}_{ab}\xi^{(0)b},
\end{eqnarray}
%
and for the source part, 
%
\begin{eqnarray}
& & \xi^{(3),\text{sou}}_{\eta}=-\frac{2}{3}\frac{mA^{(2)}\coth{\rho}}{\sinh{\rho}},\\
& & \xi^{(3),\text{sou}}_{\rho}=\frac{2}{3}\frac{mA^{(2)}}{\sinh^{2}{\rho}},\\
& & \xi^{(3),\text{sou}}_{A}=0.
\end{eqnarray}
%
As seen in the previous section, the $(a,b)$-components in 
Eq. (\ref{3AS}) shows that the homogeneous term, $\chi^{(3),\text{hom}}$ 
can have only $l=2$ modes. For the source terms we can confirm that 
there are no restrictions, that is, the ($a,b$)-components for the source 
part $\xi^{(3), {\rm sou}}$ trivially holds. This property was assumed to be 
hold for the general cases on the homogeneous parts in Sec. III. 
See Appendix C for useful formulae to show this.

To be summarized, the solutions for the third order structure which is asymptotically 
stationary at the third order is
%
\begin{eqnarray}
& & F^{(3)}=\frac{2}{3}\frac{mA^{(2)}}{\sinh^{3}\rho}-\frac{8m^{3}}{\sinh\rho}\\
& & \beta_{\rho}^{(3),\text{sou}}=-8m^{3}\frac{\coth{\rho}}{\sinh\rho}\\
& & h^{(3)}_{ab}=
\begin{pmatrix}
\frac{2}{3}\frac{mA^{(2)}}{\sinh^{3}\rho}
+8m^{3}\frac{\coth^{2}\rho}{\sinh^{3}\rho} &0 \\
0&-\frac{4}{9}\frac{mA^{(2)}}{\sinh{\rho}}\sigma_{AB}
\end{pmatrix}
+\chi^{(3),\text{hom}}_{ab},
\end{eqnarray}
%
where $\chi^{(3),\text{hom}}_{ab}$ has only $l=2$ modes
%
\begin{eqnarray}
\chi^{(3),\text{hom}}_{ab}=
\begin{pmatrix}
\frac{A^{(3)}}{\sinh^{5}{\rho}} &
-\frac{1}{3}\frac{\coth{\rho}}{\sinh^{3}{\rho}}\mathcal{D}_{A}A^{(3)}
-\frac{\epsilon_{A}^{~B}\mathcal{D}_{B}B^{(3)}}{\sinh^{3}{\rho}} \\
-\frac{1}{3}\frac{\coth{\rho}}{\sinh^{3}{\rho}}\mathcal{D}_{A}A^{(3)}
-\frac{\epsilon_{A}^{~B}\mathcal{D}_{B}B^{(3)}}{\sinh^{3}{\rho}} &
\frac{1}{4}\frac{\cosh{2\rho}-2}{\sinh^{3}{\rho}}A^{(3)}\sigma_{AB}
+\frac{1}{12}\frac{\cosh{2\rho}}{\sinh^{3}{\rho}}\mathcal{D}_{A}\mathcal{D}_{B}A^{(3)}
+\frac{1}{2}\frac{\coth{\rho}}{\sinh{\rho}}\epsilon_{(A}^{~~C}
\mathcal{D}_{B)}\mathcal{D}_{C}B^{(3)}
\end{pmatrix},
\end{eqnarray}
%
and
%
\begin{gather}
A^{(3)}=a_{xx}\sin^{2}{\theta}\cos{2\phi}+a_{xy}\sin^{2}\theta\sin{2\phi}+
a_{xz}\sin\theta\cos\theta\cos\phi +a_{yz}\sin\theta\cos\theta\sin\phi
+a_{zz}(3\cos^{2}\theta -1), \\
B^{(3)}=b_{xx}\sin^{2}{\theta}\cos{2\phi}+b_{xy}\sin^{2}\theta\sin{2\phi}+
b_{xz}\sin\theta\cos\theta\cos\phi +b_{yz}\sin\theta\cos\theta\sin\phi
+b_{zz}(3\cos^{2}\theta -1).
\end{gather}
%

\section{Comparison to the Kerr spacetimes}

In this section, we will compute the multipole moments of the spacetimes 
discussed in previous section. Then we see that the physical meaning of 
the parameters contained in the second and third order structures will be clear. 
We also compare them with those of the Kerr spacetimes. See appendix D for 
the definition of the multipole moments. 

\subsection{Multipole moments up to third order}

We will compute the multipole moments of the spacetimes obtained in previous 
section. To make the calculations simple, we first perform the gauge 
transformation $x^{\mu}\rightarrow x^{\mu} +k^{\mu}$ with
%
\begin{eqnarray}
k_{\mu}\,=\,\Omega k^{(3)}_{\mu}
\end{eqnarray}
%
where
%
\begin{eqnarray}
& & k^{(3)}_{\eta}=\frac{1}{8}\frac{A^{(3)}}{\sinh^{3}{\rho}},\\
& & k^{(3)}_{\rho}=\frac{1}{8}\frac{\cosh{\rho}}{\sinh^{4}{\rho}}A^{(3)}, \\
& & k^{(3)}_{A}=-\frac{1}{24}\frac{\cosh^{2}{\rho}}{\sinh^{3}{\rho}}\mathcal{D}_{A}A^{(3)}
-\frac{1}{4}\frac{\cosh{\rho}}{\sinh^{2}{\rho}}\epsilon_{A}^{~B}D_{B}B^{(3)}.
\end{eqnarray}
%
In this gauge, as stated later again, the asymptotic Killing vector becomes to be 
$-\partial_t$ up to the third orders. 
Then metric function of the third order structure are transformed to
%
\begin{eqnarray}
& & F^{(3)}=-8m^{3}-\frac{2}{3}\frac{mA^{(2)}}{\sinh^{3}{\rho}}
-\frac{1}{2}\frac{A^{(3)}}{\sinh^{3}{\rho}}, \\
& & \beta^{(3)}_{\rho}\,=\,-8m^{3}\frac{\coth{\rho}}{\sinh{\rho}},\\
& & \beta^{(3)}_{A}\,=\,-\frac{1}{8}\frac{\mathcal{D}_{A}A^{(3)}}{\sinh{\rho}}
-\frac{3}{4}\frac{\coth{\rho}}{\sinh{\rho}}\epsilon_{A}^{~B}\mathcal{D}_{B}B^{(3)}, \\
& & h^{(3)}_{\rho\rho}=\frac{4m^{3}}{\sinh^{3}{\rho}}+\frac{2}{3}\frac{mA^{(3)}}{\sinh^{3}{\rho}}
-\frac{1}{2}\frac{A^{(3)}}{\sinh^{3}{\rho}}, \\
& & h^{(3)}_{\rho A}=\frac{1}{8}\frac{\coth{\rho}}{\sinh{\rho}}\mathcal{D}_{A}A^{(3)}
+\frac{3}{4}\frac{\coth{\rho}}{\sinh{\rho}}\epsilon _{A}^{~B}\mathcal{D}_{B}B^{(3)},\\
& & h^{(3)}_{AB}=\left(
\frac{A^{(3)}}{\sinh{\rho}}-\frac{4}{9}\frac{mA^{(2)}}{\sinh{\rho}}
\right)\sigma_{AB}+\frac{1}{12}\frac{\mathcal{D}_{A}\mathcal{D}_{B}A^{(3)}}{\sinh{\rho}}.
\end{eqnarray}
%
Since we performed the third order gauge transformations, the first and second 
order structures do not change. 
Next we perform the coordinate transformation introduced by 
%
\begin{eqnarray}
& & t=\Omega^{-1}\cosh{\rho} \\
& & r=\Omega^{-1}\sinh{\rho}.
\end{eqnarray}
%
The transformed metric can be written as
%
\begin{eqnarray}
g_{tt}&=&
-1+\frac{2m}{r}+\frac{A^{(2)}}{4r^2}-\frac{1}{r^3}\left( \frac{2mA^{(2)}}{3}+\frac{A^{(3)}}{2}
\right) +O(\Omega^{4}), \\
g_{rr}&=&
1+\frac{2m}{r}+\frac{1}{r^2}\left( 4m^2-\frac{A^{(2)}}{3}\right) +
\frac{1}{r^3}\left( 8m^{3}+\frac{2mA^{(2)}}{3}
-\frac{A^{(3)}}{2}\right) +O(\Omega^{4}), \\
g_{tA}&=&
\frac{2}{3r}\mathcal\epsilon_{A}^{B}{D}_{B}B^{(2)}
+\frac{3}{4r^{2}}\mathcal\epsilon_{A}^{B}{D}_{B}B^{(3)} +O(\Omega^{3}),\\
g_{rA}&=&
\frac{1}{6r}\mathcal{D}_{A}A^{(2)}+\frac{1}{8r^{2}}\mathcal{D}_{A}A^{(3)} +O(\Omega^{3}), \\
g_{AB}&=&
r^{2}\sigma_{AB}+\frac{5A^{(2)}}{6}\sigma_{AB}+\frac{1}{r}
\Big[\left( A^{(3)}-\frac{4mA^{(2)}}{9}\right)
+\frac{\mathcal{D}_{A}\mathcal{D}_{B}A^{(3)}}{12} 
\Big] +O(\Omega^{2}) ,
\end{eqnarray}
%
and $g_{tr}=0$. Here $O(\Omega^{4})$ means $O(1/t^{4})$. 
In this coordinate, note that the timelike Killing vector is 
$\xi=-\partial/\partial t$ as the Kerr spacetime. 
We can obtain $\lambda=-g_{tt}$ and the twist function 
$\omega$ as (See Eq. (\ref{omega}) for the definition)
%
\begin{eqnarray}
\omega\,=\,
\frac{2B^{(2)}}{3r^{2}}+\frac{3B^{(3)}}{2r^{3}} +O(1/t^{4}).
\end{eqnarray}
%
We choose $\Omega=1/r^2$, which satisfies the conditions of Eq. (\ref{omegacond}) 
in the definition of spatial infinity. Introducing 
the coordinate $R$ defined by $R=1/r$, 
the metric $h_{\mu\nu}=\lambda g_{\mu\nu}+\xi_{\mu}\xi_{\nu}$ 
can be computed as
%
\begin{eqnarray}
h_{RR}&=&
1-\frac{2}{3}A^{(2)}R^{2}-\left( A^{(3)}-\frac{4}{3}mA^{(2)}\right) R^{3}
+O(1/t^{4}), \\
h_{R A}&=&
-\frac{\mathcal{D}_{A}A^{2}}{6}R^{3}-\left(\frac{\mathcal{D}_{A}A^{(3)}}{8}-
\frac{m\mathcal{D}_{A}A^{(2)}}{3}\right) R^{4} +O(1/t^{5}),\\
h_{AB}&=&
R^{2}\sigma_{AB}+\frac{1}{2}A^{(2)}R^{4}\sigma_{AB}
+\left(\frac{1}{3}A^{(3)}-\frac{13}{9}mA^{(2)}\right)
R^{5}\sigma_{AB}+\mathcal{D}_{A}\mathcal{D}_{B}A^{(3)}R^{5} +O(1/t^{6}).
\end{eqnarray}
%

Now we can compute the multipole moments defined in Appendix D. 
There are two types, that is, mass and spin multipole moments. 
The results are as follows. The monopole moments are
%
\begin{eqnarray}
& & P^{M}\tilde{=}-m,\\
& & P^{J}\tilde{=}0,
\end{eqnarray}
%
where $\tilde{=}$ denotes the evaluation at $r=\infty$. The dipole moments are 
%
\begin{eqnarray}
& & P^{M}_{R}\tilde{=}-\frac{1}{6}A^{(2)}-m^2 ,\\
& & P^{J}_{R}\tilde{=}\frac{1}{3}B^{(2)}.
\end{eqnarray}
%
In the mass dipole moment, the second terms comes from the $l=0$ mode of scalar harmonics 
on sphere while the first terms does from the $l=1$ mode. The spin 
dipole moment has only the contribution from the $l=1$ mode of scalar harmonics. 
The quadrapole moments are 
%
\begin{eqnarray}
& & P^{M}_{RR}\tilde{=}-\frac{1}{2}A^{(3)}-\frac{2}{3}mA^{(2)}-2m^{3} ,\\
& & P^{J}_{RR}\tilde{=}\frac{3}{2}B^{(3)}+\frac{3}{2}mB^{(2)}.
\end{eqnarray}
%
The first term in the both comes from $l=2$ modes. Other terms are contribution from 
the $l=0$ or $l=1$ modes. 

From the above results, we observe that the second and third order structures 
have the information of the dipole and quadrapole moments of spacetimes. 
In particular, the even parity modes $A^{(2)}$, $A^{(3)}$ 
correspond to the mass multipole moments and the odd parity modes 
$B^{(2)}$, $B^{(3)}$ correspond to 
the spin multipole moments. We would guess that this property might be valid at $n$-th order 
and this implies that stationary space-time can be determined completely by 
the mass and spin multipole moments. 

In the Kerr spacetime, non-trivial multipole moments are only mass monopole and 
$m=0$ mode of spin dipole moment. Other multipole moments, e.g., 
mass quadrupole moments, $a_{zz}$ in $A^{(3)}$,  
can be written by mass monopole and spin dipole moments as $a_{zz}=Ma^{2}$ where $M$ 
is the mass of the Kerr 
black hole and $a$ is the Kerr parameter. However, the spacetime we obtained has not 
only nontrivial mass quadrupole moments, that is, $a_{zz}$ is a free parameter, 
but also $l\neq 0$ modes of 
multipole moments, e.g., $a_{xx}$, $a_{xy}$, $a_{xz}$ and $a_{yz}$ in $A^{(3)}$. 
This means that asymptotically stationary spacetimes can deviate from the Kerr spacetime.

\section{summary and discussion}
\label{Sec:SD}

In this paper we study higher order asymptotic structure at timelike infinity. 
There are two tensor modes as degree of freedom, that is, 
even and odd modes. 
They correspond to the degree of freedom of gravity in four dimensions. 
Imposing asymptotic stationarity at $n$-th order, we could show that 
only $l=n-1$ mode 
can be allowed. As shown in Sec. IV, in the case of second and third 
order structures, 
the even modes correspond to the mass multipole moments and 
the odd mode corresponds to the spin multipole 
moments. 
This property would hold in $n$-th order too. At first order, the degree 
of freedoms are contained only in the scalar mode which corresponds to the mass and 
the asymptotic axisymmetry is realized. On the other hand, at higher orders, we cannot 
currently guarantee the asymptotic axisymmetry only by imposing the asymptotically stationary condition. 
Hence, the late time phases of dynamical spacetimes 
will deviate from the Kerr spacetime in our approach. However these deviations contain 
unphysical and singular modes such as outgoing mode at horizon because we did not impose 
the boundary condition at event horizon. Thus, another boundary condition near the 
horizon might be needed to extract physical degree of freedom. As a first step, 
it might be better to analyze the asymptotic structures near horizons. If one focuses 
on the event horizon, we can show that the expansion and shear of the outgoing null 
geodesics congruence approaches to zero as the time goes by \cite{Hayward:1993tt}. 
If the ergoregion exists, 
this implies the another asymptotic isometry which could be decomposed into the asymptotic stationary 
and rotational isometries in the similar way with the rigidity theorem for stationary 
spacetimes \cite{Hawking:1971vc}. So we can expect the presence of asymptotic rotational symmetry near the 
event horizon and then $m \neq 0$ modes will be killed. Although we do not have the proof for the moment, this asymptotic 
rotational symmetry would present in outside of the horizon too(``asymptotic rigidity"). 
However, as seen in the previous section, this additional symmetry is not enough 
to show the asymptotic uniqueness. For this, some informations about event horizon 
topology may be important. 
These will be our future work. 

As an application of our work to higher dimensions, one may be able to construct the 
asymptotic form of stationary black objects in higher dimensions. For example, in 
five dimensions we have only black object solutions which have two rotational 
Killing vector. However it is conjectured that there are solutions which have only 
one rotational Killing vector \cite{Reall:2002bh}. Then it may be possible 
to obtain the asymptotic form of such spacetime by using our method. The asymptotic form 
may give us a new implication into the study on the founding of the exact solutions. 
It may be useful to classify the higher dimensional spacetimes. In five dimensions, 
the mass and spin multipole moments have been proposed \cite{Tanabe:2010ax}. But, it 
was turned out that these quantities are not enough to classify the spacetimes. 
The asymptotic form of stationary spacetimes in higher dimensions provides us the 
chance to look for the good parameters. 

\section*{Acknowledgment}
KT is supported by JSPS Grant-Aid for Scientific Research (No.$~21$-$2105$). 
TS is partially supported by
Grant-Aid for Scientific Research from Ministry of Education, Science,
Sports and Culture of Japan (Nos.$~21244033,~21111006,~20540258$ and $19$GS$0219$). 
This work is also supported by the Grant-in-Aid for the Global 
COE Program ``The Next Generation of Physics, Spun from Universality 
and Emergence'' from the Ministry of Education, Culture, Sports, Science 
and Technology (MEXT) of Japan.

\appendix

\section{Tensor harmonics on hyperboloid}

We introduce the tensor harmonics $\mathcal{G}^{\pm ,plm}_{ab}$ on three 
dimensional hyperboloid. See Ref. \cite{Tanaka:1997kq} for examples. The tensor harmonics satisfies 
%
\begin{eqnarray}
(D^{2}+3-(p-1)^2)\mathcal{G}^{\pm ,plm}_{ab}\,=\,0,
\end{eqnarray}
%
where $D^{2}=D^{a}D_{a}$ and $D_{a}$ is the covariant derivative on the hyperboloid. 
The metric on the hyperboloid is
%
\begin{eqnarray}
ds^{2}&=&d\rho^{2}+\sinh^{2}{\rho}(d\theta^{2}+\sin^{2}{\theta}d\phi^{2}) \notag \\
      &=&d\rho^{2}+\sinh^{2}{\rho}\sigma_{AB}dx^{A}dx^{B},
\end{eqnarray}
%
where $\sigma_{AB}$ is the metric of the two-dimensional sphere. 

Now $\mathcal{G}^{\pm ,plm}_{ab}$ can be written as
%
\begin{gather}
\mathcal{G}^{+ ,plm}_{\rho\rho}=\mathcal{T}_{1}^{pl}Y^{lm} \,,\,
\mathcal{G}^{+, plm}_{\rho A}=\mathcal{T}_{2}^{pl}Y^{lm} \notag\\
\mathcal{G}^{+, plm}_{AB}=\mathcal{T}_{3}^{pl}\mathcal{D}_{A}\mathcal{D}_{B}Y^{lm} \\
\mathcal{G}^{-, plm}_{\rho\rho}=0\,,\,
\mathcal{G}^{-, plm}_{\rho A}=\mathcal{T}_{5}^{pl}\mathcal{Y}^{lm}_{A}\,,\,
\mathcal{G}^{-, plm}_{AB}2\mathcal{T}_{6}^{pl}\mathcal{Y}^{lm}_{AB},
\end{gather}
%
where $Y^{lm}(\theta,\phi)$ are spherical harmonics on $S^{2}$ and
%
\begin{eqnarray}
\mathcal{Y}_{A}=\epsilon_{~A}^{B}{\cal D}_{B}Y^{lm}\,,\,
\mathcal{Y}_{AB}=\epsilon^{C}_{~(B}\mathcal{D}_{A)}\mathcal{D}_{C}Y^{lm},
\end{eqnarray}
%
where ${\cal D}_A$ is the  covariant derivative and 
$\epsilon_{AB}$ is the antisymmetric tensor on $S^{2}$ with 
$\epsilon_{\theta\phi}=\sin\theta$, 
The $\rho$-dependent parts of tensor harmonics are 
%
\begin{eqnarray}
\mathcal{T}^{n,l}_{1} &=& \frac{1}{\sinh^{2}{\rho}}P^{n,l}(\rho) \\
\mathcal{T}^{n,l}_{2} &=& \frac{1}{l(l+1)}(\partial_{\rho}+\coth{\rho})P^{n,l} \\
\mathcal{T}^{n,l}_{3} &=& \frac{2\sinh^{2}{\rho}}{(l-1)l(l+1)(l+2)}
\left[\coth{\rho}\partial_{\rho} +\left( n^2+1+\frac{l(l+1)+2}{2\sinh^{2}{\rho}} 
\right)\right] P^{n,l} \\
\mathcal{T}^{n,l}_{4} &=& \frac{\sinh^{2}{\rho}}{(l-1)(l+2)}\left[
\coth{\rho}\partial_{\rho} + \left( n^2+1+\frac{2}{\sinh^{2}{\rho}} 
\right)\right] P^{n,l} \\
\mathcal{T}^{n,l}_{5} &=& P^{n,l} \\
\mathcal{T}^{n,l}_{6} &=& \frac{\sinh^{2}{\rho}}{(l-1)(l+2)} 
(\partial_{\rho} + 2\coth{\rho})P^{n,l}.
\end{eqnarray}
%
$P^{n,l}$ is the function defined  as
%
\begin{gather}
P^{n,l}\,=\,\frac{1}{\sqrt{\sinh{\rho}}}\mathcal{P}^{l+1/2}_{n-1/2}(\cosh{\rho}),
\end{gather}
%
where $\mathcal{P}^{l}_{n}$ is the first kind associated Legendre function of  
and $P^{n,l}$ satisfies  
%
\begin{gather}
\partial_{\rho}^{2}P^{n,l}+2\coth{\rho}\partial_{\rho}P^{n,l}-
\left(n^2-1+\frac{l(l+1)}{\sinh^{2}{\rho}}
\right)P^{n,l}
\,=\,0.
\end{gather}
%

\section{Zero-th and first order structures}

Here we discuss the zero-th and first order structure. See 
Ref. \cite{Gen:1997az, Gen:1998ay} for the details. 

\subsection{Zero-th order structure}

We can write the metric as
%
\begin{eqnarray}
g_{\mu\nu}\,&=&\,\Omega^{-2}[-\Omega^{-2}F^{-1}\nabla_{\mu}\Omega\nabla_{\nu}\Omega
+q_{\mu\nu}] \notag \\
            &=&\, -\hat{n}_{\mu}\hat{n}_{\nu}+\hat{q}_{\mu\nu},
\end{eqnarray}
%
where $\hat{n}_{\mu}=\Omega^{-2}F^{-1/2}\nabla_{\mu}\Omega$ is the normal vector of 
$\Omega=\text{const.}$ hypersurface and $\hat{q}_{\mu\nu}=\Omega^{-2}q_{\mu\nu}$ 
is the metric on its hypersurface. $\Omega=0$ is the timelike infinity.
The extrinsic curvature $K_{\mu\nu}$ of 
$\Omega=\text{const.}$ hypersurface is expressed as 
%
\begin{eqnarray}
\hat{K}_{\mu\nu}\,&=&\,\frac{1}{2}\mathcal{L}_{\hat{n}}\hat{q}_{\mu\nu} \notag \\
&=&\Omega^{-1}F^{1/2}q_{\mu\nu}+\frac{1}{2}F^{-1/2}\mathcal{L}_{n}q_{\mu\nu},
\label{extrinsic}
\end{eqnarray}
%
where $n^{\mu}=\Omega^{-2}F^{1/2}n^{\mu}=-(\partial\Omega)^{\mu}$. We define
$K_{\mu\nu}=\Omega\hat{K}_{\mu\nu}$. Now the Einstein equations $G_{\mu\nu}\hat n^{\mu} \hat q^{\mu}_{~\nu}=0$
can be written as
%
\begin{eqnarray}
\hat{D}_{\nu}\hat{K}^{~\nu}_{\mu}-\hat{D}_{\mu}\hat{K}=
\Omega^{-1}(D_{\nu}K^{~\nu}_{\mu}-D_{\mu}K)=0 \label{momentconst},
\end{eqnarray}
%
where $\hat{D}$ and $D$ are the covariant derivatives with respect to 
$\hat{q}_{\mu\nu}$ and $q_{\mu\nu}$, respectively. 
From Eqs. (\ref{extrinsic}) and (\ref{momentconst}) we see that 
$F\,\hat{=}\,\text{const.}$ at timelike infinity. 
Using the freedom such as $\Omega\rightarrow a\Omega$ where
$a$ is a constant, we can always take $F\hat{=}1$. 
Then $K_{\mu\nu}\hat{=}q_{\mu\nu}$ at timelike infinity.

Next, from the Einstein equation $R_{\mu\nu}\hat{q}^{\mu}_{~\rho}\hat{q}^{\nu}_{~\sigma}=0$, 
we can see that $q_{\mu\nu}$ satisfies the following equation
%
\begin{eqnarray}
\,^{(3)}R_{\mu\nu}+KK_{\mu\nu}+F^{1/2}K_{\mu\nu}-2K_{\mu\rho}K^{\rho}_{~\nu}
+\Omega F^{-1/2}\mathcal{L}_{n}K_{\mu\nu}+F^{-1/2}D_{\mu}D_{\nu}F^{1/2}=0,
\end{eqnarray}
%
where $\,^{(3)}R_{\mu\nu}$ is the Ricci tensor of $q_{\mu\nu}$.
At timelike infinity, this equation becomes $\,^{(3)}R_{\mu\nu}\hat{=}-2q_{\mu\nu}$. Thus 
$q_{\mu\nu}$ should be the metric of the unit-hyperboloid at timelike infinity
%
\begin{eqnarray}
q_{\mu\nu}dx^{\mu}dx^{\nu}\,\hat{=}\,d\rho^{2}+\sinh^{2}\rho(d\theta^{2}+\sin^{2}\theta d\phi^{2}).
\end{eqnarray}
%
To be summarized, from Einstein equation we obtained the zero-th order structure such as
%
\begin{eqnarray}
g_{\mu\nu}dx^{\mu}dx^{\nu}&\hat{=}&\Omega^{-2}[-\Omega^{-2}d\Omega^2+
d\rho^{2}+\sinh^{2}\rho(d\theta^{2}+\sin^{2}\theta d\phi^{2})] \notag \\
&\hat{=}&e^{-2\eta}[-d\eta^2+
h^{(0)}_{ab}dx^{a}dx^{b}],
\end{eqnarray}
%
where $\Omega=e^{\eta}$ and $x^{a}=(\rho, \theta, \phi)$. The timelike infinity is located 
at $\eta=-\infty$. 

\subsection{First order structure}

We solve the Einstein equation for the first order structure $g^{(1)}_{\mu\nu}$
%
\begin{eqnarray}
g^{(1)}_{\mu\nu}=e^{-2\eta}[-F^{(1)}d\eta^{2}+2\beta^{(1)}_{a}dx^{a}d\eta +h^{(1)}_{ab}dx^{a}dx^{b}],
\end{eqnarray}
%
where $h^{(1)}_{ab}=\psi^{(1)}h^{(0)}_{ab}+\chi^{(1)}_{ab}$. As shown in Sec. III, taking 
the Poisson gauge ($\beta^{(1)}_{a}=0, \psi^{(1)}=-F^{(1)}$), the 
Einstein equations become
\footnote{Since there are no source term at first order structure, we should solve the 
homogeneous part only.}
%
\begin{eqnarray}
(D^{2}-3)F^{(1)}\,=\,0 \,,\,
(D^{2}+3)\chi^{(1)}_{ab}\,=\,0 \label{1steq}
\end{eqnarray}
%
As seen in Sec. III, we can always take $F^{(n)}=0$ by residual gauge for $n>1$. But, 
we cannot do it at first order because $F^{(1)}$ is gauge invariant quantity. 
Instead, we can eliminate the even parity 
mode $\chi^{+,(1)}_{ab}$ by residual gauges. Then, the solutions to Eq. (\ref{1steq}) are
%
\begin{eqnarray}
F^{(1)}\,=\,\sum_{l.m}a^{(1)}_{l,m}P^{2,l}Y^{lm} \\
\chi^{(1)}_{ab}\,=\,\sum_{l,m}b^{(1)}_{l,m}\mathcal{G}^{-,1lm}.
\end{eqnarray}
%
Next we impose the asymptotic stationary condition. This condition is
%
\begin{gather}
\nabla_{\mu}\xi_{\nu}+\nabla_{\nu}\xi_{\mu}\,=\,O(\Omega) \label{1stkill},
\end{gather}
%
where $\xi_{\mu}=\Omega^{-1}\xi^{(0)}_{\mu}+\xi_{\mu}$ is an asymptotic Killing vector. 
From the $(\eta,\mu)$ components  of Eq. (\ref{1stkill}), we can see
%
\begin{eqnarray}
\xi^{(1)}_{\eta}&=&-\frac{1}{2}(\cosh\rho F^{(1)}-\sinh\rho D_{\rho}F^{(1)}) \\
\xi^{(1)}_{a} &=&-\frac{1}{4}(\sinh\rho D_{a}D_{\rho}F^{(1)}+2\cosh\rho D_{\rho}F^{(1)}
+F^{(1)}\xi^{(0)}_{a}-2\chi^{(1)}_{ab}\xi^{(0)b}).
\end{eqnarray}
%
Then the $(a,b)$-components of Eq. (\ref{1stkill}) gives us 
%
\begin{eqnarray}
\frac{3}{\tanh\rho}(D_{a}D_{b}-h^{(0)}_{ab})F^{(1)}+D_{a}D_{b}D_{\rho}F^{(1)} -
h^{(0)}_{\rho (a}D_{b)}F^{(1)}+2D_{(a}\chi^{(2)}_{b)\rho} -2D_{\rho}\chi^{(1)}_{ab}=0.
\end{eqnarray}
%
This equation is satisfied if and only if $\chi^{(1)}_{ab}=0$ and $F^{(1)}$ has 
only $l=0$ mode. From the fact $P^{20}=\cosh 2\rho/\sinh\rho$, the asymptotically stationary 
first order structure is
%
\begin{eqnarray}
(g^{(0)}_{\mu\nu}+\Omega g^{(1)}_{\mu\nu})dx^{\mu}dx^{\nu}
=e^{-2\eta}\Big[-(1-2m\frac{\cosh 2\rho}{\sinh\rho}\Omega)d\eta^{2}
+(1+2m\frac{\cosh 2\rho}{\sinh\rho}\Omega)h^{(0)}_{ab}dx^{a}dx^{b}
\Big],
\end{eqnarray}
%
where we define $a^{(1)}_{00}=-2m$. Finally we perform the gauge transformation 
$x^{\mu}\rightarrow x^{\mu}+k^{\mu}$ as 
%
\begin{eqnarray}
k_{\eta}\,=\,2m(2\rho\cosh\rho +\sinh\rho)\,,\,
k_{\rho}\,=\,2m(2\rho\sinh\rho -\cosh\rho),
\end{eqnarray}
%
and other components are zero. Then, the metric is transformed to
%
\begin{eqnarray}
(g^{(0)}_{\mu\nu}+\Omega g^{(1)}_{\mu\nu})dx^{\mu}dx^{\nu}\,&=&\,e^{-2\eta}
\left[
(-1+2m\frac{\cosh{2\rho}}{\sinh{\rho}}\Omega ) d\eta^{2}-8m\cosh{\rho}\Omega d\eta d\rho
\right. 
\notag \\
&&~~~~~~~~~~~~~~~~~~~~~~~~~~~~~~~~~
+(1+2m\frac{\cosh{2\rho}}{\sinh{\rho}}\Omega ) d\rho^{2} 
+\sinh^{2}{\rho}(d\theta^{2}+\sin^{2}{\theta}d\phi^{2})
\bigg].
\end{eqnarray}
%
There is an asymptotic Killing vector up to the first order as
%
\begin{eqnarray}
\xi_{\mu}\,=\,\Omega^{-1}\xi^{(0)}_{\mu} + \xi_{\mu}^{(1)},
\end{eqnarray}
%
where
%
\begin{eqnarray}
\xi^{(1)}\,=\,2m\coth{\rho}d\eta -2m d\rho.
\end{eqnarray}
%

\section{Some useful formulae for second and third order structures}

\subsection{for second order}

The metric and its dual are expanded as 
%
\begin{eqnarray}
ds^{2}\,=\,[g^{(0)}_{\mu\nu}+\Omega g^{(1)}_{\mu\nu}+\Omega^{2}g^{(2)}_{\mu\nu}
]dx^{\mu}dx^{\nu}
\end{eqnarray}
%
and
%
\begin{eqnarray}
g^{\mu\nu}\partial_\mu \partial_\nu
=\Bigl[ g^{(0)\mu\nu}+g^{(1)\mu\nu}\Omega+g^{(2)\mu\nu}\Omega^2+\cdots \Bigr]\partial_\mu \partial_\nu.
\end{eqnarray}
%
In each orders, 
%
\begin{eqnarray}
g^{(0)}_{\mu\nu}dx^{\mu}dx^{\nu}\,=\,e^{-2\eta}[-d\eta^{2}+d\rho^{2}+\sinh^{2}{\rho}\sigma_{AB}
dx^{A}dx^{B}
], \label{metric0}
\end{eqnarray}
%
%
\begin{eqnarray}
g^{(1)}_{\mu\nu}dx^{\mu}dx^{\nu}\,=\,e^{-2\eta}\Bigg[
2m\frac{\cosh{2\rho}}{\sinh{\rho}}d\eta^{2}-8m\cosh{\rho}d\eta d\rho
+2m\frac{\cosh{2\rho}}{\sinh{\rho}}d\rho^{2}
\Bigg], \label{metric1}
\end{eqnarray}
%
%
\begin{eqnarray}
g^{(2), {\rm sou}}_{\mu\nu}dx^{\mu}dx^{\nu}\,=\,e^{-2\eta}\Bigg[
 4m^{2} d\eta^{2}-8m^{2}\coth{\rho}d\eta d\rho
+4m^{2}\coth^{2} d\rho^{2}
\Bigg].
\end{eqnarray}
%
%
\begin{eqnarray}
g^{(0)\mu\nu}\partial_{\mu}\partial_{\nu}\,=\,e^{2\eta}[-\partial_{\eta}^{2}
+\partial_{\rho}^{2}+(\sinh^{2}{\rho})^{-1}\sigma^{AB}
\partial_{A}\partial_{B}] \label{dualmetric0}
\end{eqnarray}
%
%
\begin{eqnarray}
g^{(1)\mu\nu}\partial_{\mu}\partial_{\nu}\,=\,e^{2\eta}
\Bigg[
-2m\frac{\cosh{2\rho}}{\sinh{\rho}}\partial^{2}_{\eta}-8m\cosh{\rho}\partial_{\eta}\partial_{\rho}
-2m\frac{\cosh{2\rho}}{\sinh{\rho}}\partial_{\rho}^{2}
\Bigg] \label{dualmetric1}
\end{eqnarray}
%
and
%
\begin{eqnarray}
g^{(2)\mu\nu, {\rm sou}}\partial_{\mu}\partial_{\nu}\,=\,e^{2\eta}
\Bigg[
-4m^{2}\coth^{2}{\rho}\partial^{2}_{\eta}-8m^{2}\coth{\rho}\partial_{\eta}\partial_{\rho}
-4m^{2}\partial_{\rho}^{2}
\Bigg].
\end{eqnarray}
%
The affine connection $\Gamma^\alpha_{\mu\nu}$ is expanded as 
%
\begin{eqnarray}
\Gamma^{\alpha}_{\mu\nu}
={}^{(0)}\Gamma^{\alpha}_{\mu\nu}+{}^{(1)}\Gamma^{\alpha}_{\mu\nu}
+{}^{(2)}\Gamma^{\alpha}_{\mu\nu}+\cdots
\end{eqnarray}
%
and the each components become 
%
\begin{eqnarray}
{}^{(1)}\Gamma^{\eta}_{\rho\rho}\,=\,-m\frac{3\cosh{2\rho}-4\sinh^{2}{\rho}}{\sinh{\rho}}
\,,\,
{}^{(2)}\Gamma^{\eta, {\rm sou}}_{\rho\rho}\,=\,-2m^{2}\frac{3+4\sinh^{2}{\rho}}{\sinh^{2}{\rho}},
\end{eqnarray}
%
%
\begin{eqnarray}
{}^{(1)}\Gamma^{\rho}_{\rho\rho}\,=\,-m\frac{\cosh{\rho}\cosh{2\rho}}{\sinh^{2}{\rho}}
\,,\,
{}^{(2)}\Gamma^{\rho, {\rm sou}}_{\rho\rho}\,=\,-2m^{2}\frac{\cosh{\rho}(1+4\sinh^{2}{\rho})}{\sinh^{3}{\rho}},
\end{eqnarray}
%
%
\begin{eqnarray}
{}^{(1)}\Gamma^{\eta}_{\rho A}\,=\,0
\,,\,
{}^{(2)}\Gamma^{\eta, {\rm sou}}_{\rho A}\,=\,0,
\end{eqnarray}
%
%
\begin{eqnarray}
{}^{(1)}\Gamma^{\rho}_{\rho A}\,=\,0
\,,\,
{}^{(2)}\Gamma^{\rho, {\rm sou}}_{\rho A}\,=\,0,
\end{eqnarray}
%
%
\begin{eqnarray}
{}^{(1)}\Gamma^{\eta}_{AB}\,=\,2m\sinh{\rho}\sigma_{AB}
\,,\,
{}^{(2)}\Gamma^{\eta, {\rm sou}}_{AB}\,=\,0,
\end{eqnarray}
%
%
\begin{eqnarray}
{}^{(1)}\Gamma^{\rho}_{AB}\,=\,2m\cosh{\rho}\sigma_{AB}
\,,\,
{}^{(2)}\Gamma^{\rho, {\rm sou}}_{AB}\,=\,0.
\end{eqnarray}
%
Here note that we focused on the ``source" parts in the second order quantities.

\subsection{for third order}

The metric and its dual are expanded up to the third order as
%
\begin{eqnarray}
ds^{2}\,=\,[g^{(0)}_{\mu\nu}+\Omega g^{(1)}_{\mu\nu}+\Omega^{2}g^{(2)}_{\mu\nu}
+\Omega^{3}g^{(3)}_{\mu\nu}]dx^{\mu}dx^{\nu}
\end{eqnarray}
%
and
%
\begin{eqnarray}
ds^{2}\,=\,[g^{(0)\mu\nu}+\Omega g^{(1)\mu\nu}+\Omega^{2}g^{(2)\mu\nu}
+\Omega^{3}g^{(3)\mu\nu}]\partial_{\mu}\partial_{\nu}
\end{eqnarray}
%
where $g^{(0)}_{\mu\nu}$, $g^{(0)\mu\nu}$, $g^{(1)}_{\mu\nu}$ and $g^{(1)\mu\nu}$ 
are given in Eqs. (\ref{metric0}), (\ref{metric1}), (\ref{dualmetric0}) and (\ref{dualmetric1}). 

Taking account of the gauge transformations given in the text,  
$g^{(2)}_{\mu\nu}$, $g^{(3), {\rm sou}}_{\mu\nu}$ and 
$g^{(2){\mu\nu}}$ and $g^{(3)\mu\nu, {\rm sou}}$ are 
%
\begin{eqnarray}
g^{(2)}_{\mu\nu}dx^{\mu}dx^{\nu}&=& e^{-2\eta}\Bigg[
\left( 4m^{2}+\frac{A^{(2)}}{3\sinh^{2}{\rho}}\right)d\eta^{2}-8m^{2}\coth{\rho}d\eta d\rho
\notag \\
&&~~~~~~~~~
-\frac{1}{3}(\mathcal{D}_{A}A^{(2)}+4\coth{\rho}\epsilon_{AB}\mathcal{D}^{B}B^{(2)})dx^{A}d\eta
+\left( 4m^{2}\coth^{2}{\rho}-\frac{A^{(2)}}{3\sinh^{2}{\rho}}\right) d\rho^{2}
\notag \\
&&~~~~~~~~~~~~~~~~~~~~~~~~~~
+\frac{1}{3}(\coth{\rho}\mathcal{D}_{A}A^{(2)}+4\epsilon_{AB}\mathcal{D}^{B}B^{(2)})dx^{A}d\rho
+\frac{5}{6}A^{(2)}\sigma_{AB}dx^{A}dx^{B}
\Bigg],
\end{eqnarray}
%
%
\begin{eqnarray}
g^{(3), {\rm sou}}_{\mu\nu}dx^{\mu}dx^{\nu}&=& e^{-2\eta}\Bigg[
\left( \frac{8m^{3}}{\sinh{\rho}}-\frac{2mA^{(2)}}{3\sinh^{3}{\rho}} \right) d\eta^{2}
-16m^{3}\frac{\coth{\rho}}{\sinh{\rho}}d\eta d\rho
\notag \\
&&~~~~~~~~~~~~~~~~~~~~~~~~~~~~~~
+\left( 8m^{3}\frac{\coth^{2}{\rho}}{\sinh{\rho}}+\frac{2mA^{(2)}}{3\sinh^{3}{\rho}} \right) d\rho^{2}
-\frac{4mA^{(2)}}{9\sinh{\rho}}\sigma_{AB}dx^{A}dx^{B}
\Bigg],
\end{eqnarray}
%
and
%
\begin{eqnarray}
g^{(2)\mu\nu}\partial_{\mu\nu}&=&e^{2\eta}\Bigg[
-\left( 4m^{2}\coth^{2}{\rho}+\frac{A^{(2)}}{3\sinh^{2}{\rho}} \right)\partial^{2}_{\eta}
-8m^{2}\coth{\rho}\partial_{\eta}\partial_{\rho}
\notag \\
&&~~~~~~~~~~~~
-\frac{1}{3\sinh^{2}{\rho}}\left( \mathcal{D}^{A}A^{(2)}+4\coth{\rho}\epsilon^{AB}
\mathcal{D}_{B}B^{(2)} \right)\partial_{A}\partial_{\eta}
-\left( 4m^{2}-\frac{A^{(2)}}{3\sinh^{2}{\rho}} \right)\partial^{2}_{\rho}
\notag \\
&&~~~~~~~~~~~~~~
-\frac{1}{3\sinh^{2}{\rho}}\left( \coth{\rho}\mathcal{D}^{A}A^{(2)}+4\epsilon^{AB}
\mathcal{D}_{B}B^{(2)} \right)\partial_{A}\partial_{\rho}
-\frac{5}{6}\frac{A^{(2)}}{\sinh^{4}{\rho}}\sigma^{AB}\partial_{A}\partial_{B}
\Bigg] \\
g^{(3)\mu\nu, {\rm sou}}\partial_{\mu\nu}&=&e^{2\eta}\Bigg[
-\frac{m}{\sinh{\rho}}\left( 8m^{2}\coth^{2}{\rho}+\frac{2A^{(2)}}{3}
\frac{2\cosh{2\rho}-1}{\sinh^{2}{\rho}} \right)\partial_{\eta}^{2}
-\frac{16m}{3}(3m^{2}+A^{(2)})\frac{\coth{\rho}}{\sinh{\rho}}\partial_{\rho}\partial_{\eta}
\notag \\
&&~~~~~~~
+\frac{2m}{3\sinh^{3}{\rho}}\left( \mathcal{D}^{A}A^{(2)}-4\coth{\rho}\epsilon^{AB}
\mathcal{D}_{B}B^{(2)} \right)\partial_{A}\partial_{\eta}
-\frac{m}{\sinh{\rho}}\left( 8m^{2}+\frac{2A^{(2)}}{3}\left( 4+\frac{3}{\sinh^{2}{\rho}}\right)\right)
\notag \\
&&~~~~~~~~~~~~~~~~
+\frac{2m}{3\sinh^{3}{\rho}}\left( \coth{\rho}\mathcal{D}^{A}A^{(2)}-4\epsilon^{AB}
\mathcal{D}_{B}B^{(2)} \right)\partial_{A}\partial_{\eta}
+\frac{4}{9}\frac{mA^{(2)}}{\sinh^{5}{\rho}}\sigma^{AB}\partial_{A}\partial_{B}
\Bigg].
\end{eqnarray}
%

Then affine connections $\Gamma$ can be computed up to the third order
(only source parts for third orders)
%
\begin{eqnarray}
\Gamma^{\eta}_{\eta\eta}&=&-1-\Omega m\frac{\cosh{2\rho}}{\sinh{\rho}}
-\frac{\Omega^{2}}{3\sinh^{2}{\rho}}(6m^{2}(2\cosh{2\rho}+1)+A^{(2)})
\notag \\
&&~~~~~~~~~~~~~~~~~~~~~~~~~~~~~~~~~~
-\frac{m\Omega^{3}}{3\sinh^{3}{\rho}}(6m^{2}(1+3\cosh{2\rho}))+A{(2)}(5\cosh{2\rho}-1),
\\
\Gamma^{\rho}_{\eta\eta}&=&-m\Omega\frac{\coth{2\rho}(\cosh{2\rho}-2)}{\sinh{\rho}}
+\Omega^{2}\frac{\coth\rho}{\sinh^{2}{\rho}}\left( 2m^{2}(1-2\cosh{2\rho})+\frac{A^{(2)}}{3}\right)
\notag \\
&&~~~~~~~~~~~~~~~~~~~~~~~~~~~~~~~~~~~~~~~~~~~~
-\frac{2m\Omega^{3}}{3}\frac{\coth{\rho}}{\sinh{\rho}}
\left( 18m^{2}+A^{(2)}\left( 5+\frac{2}{\sinh^{2}{\rho}}\right)\right),
\\
\Gamma^{A}_{\eta\eta}&=&-\Omega^{2}\frac{\coth{\rho}}{6\sinh^{2}{\rho}}\left(\coth{\rho}
\mathcal{D}^{A}A^{(2)}+4\epsilon^{AB}\mathcal{D}_{B}B^{(2)}\right)
\notag \\
&&~~~~~~~~~~~~~~~~~~~~~~~~~~~~~~~~~~~~~~
+\frac{m\Omega^{3}}{6\sinh^{3}{\rho}}\left(\frac{1}{\sinh^{2}{\rho}}\mathcal{D}^{A}A^{(2)}
-8\coth{\rho}\epsilon^{AB}\mathcal{D}_{B}B^{(2)}\right),
\\
\Gamma^{\eta}_{\eta\rho}&=&\Omega m\frac{\coth{\rho}\cosh{2\rho}}{\sinh{\rho}}
+\Omega^{2}\frac{\coth{\rho}}{\sinh^{2}{\rho}}\left( 2m^{2}(2\coth{2\rho}-1)+\frac{A^{(2)}}{3}\right)
\notag \\
&&~~~~~~~~~~~~~~~~~~~~~~~~~~~~~~~~~~~~~~~~~~~~~
+\Omega^{3}m\frac{\coth{\rho}}{\sinh{\rho}}\left( 4m^{2}\left( 3+\frac{1}{\sinh^{2}{\rho}}\right)
+\frac{10}{3}A^{(2)}\right),
\\
\Gamma^{\rho}_{\eta\rho}&=&-1+\Omega m\frac{\cosh{2\rho}}{\sinh{\rho}}
+\frac{\Omega^{2}}{\sinh^{2}{\rho}}\left( 2m^{2}(2\cosh{2\rho}-1)-\frac{A^{(2)}}{3}\right)
\notag \\
&&~~~~~~~~~~~~~~~~~~~~~~~~~~~~~~~~~~~~~~~~~~~~~
+\frac{2m\Omega^{3}}{3\sinh{\rho}}\left( 3m^{2}\frac{1+3\cosh{2\rho}}{\sinh^{2}{\rho}}
+5A^{(2)}\coth^{2}{\rho}\right),
\\
\Gamma^{A}_{\eta\rho}&=&\frac{\Omega^{2}}{6\sinh^{2}{\rho}}\left( \coth{\rho}\mathcal{D}^{A}A^{(2)}
+\frac{2\cosh{2\rho}}{\sinh^{2}{\rho}}\epsilon^{AB}\mathcal{D}_{B}B^{(2)}\right)
+\frac{2m\Omega^{3}}{3\sinh^{5}{\rho}}\cosh{2\rho}\epsilon^{AB}\mathcal{D}_{B}B^{(2)},
\\
\Gamma^{\eta}_{\eta A}&=&\frac{\Omega^{2}}{12}\left(\frac{\cosh{2\rho}-3}{\sinh^{2}{\rho}}
\mathcal{D}_{A}A^{(2)}+8\coth{\rho}\epsilon_{AB}\mathcal{D}^{B}B^{(2)}\right)
\notag \\
&&~~~~~~~~~~~~~~~~~~~~~~~~~~~~~~~~~~~~~~~~~~~~~~~~~~
-\frac{m\Omega^{3}}{\sinh{\rho}}\left(\mathcal{D}_{A}A^{(2)}-\frac{8}{3}\coth{\rho}
\epsilon_{AB}\mathcal{D}^{B}B^{(2)}\right),
\end{eqnarray}
%
%
\begin{eqnarray}
\Gamma^{\rho}_{\eta A}&=&\frac{\Omega^{2}}{6}\left(\coth{\rho}\mathcal{D}_{A}A^{(2)}
-2\left(2-\frac{1}{\sinh^{2}{\rho}}\epsilon_{AB}\mathcal{D}^{B}B^{(2)}\right)\right)
\notag \\
&&~~~~~~~~~~~~~~~~~~~~~~~~~~~~~~~~~~~~~~~~~~~~~~~
-\frac{m\Omega^{3}}{\sinh{\rho}}\left(\mathcal{D}_{A}A^{(2)}
-\frac{6+2\coth^{2}{\rho}}{3}\epsilon_{AB}\mathcal{D}^{B}B^{(2)}\right),
\\
\Gamma^{A}_{\eta B}&=&-\delta^{A}_{~B}+\frac{\Omega^{2}}{\sinh^{2}{\rho}}\left(\frac{5}{6}A^{(2)}
\delta^{A}_{~B}+\frac{1}{3}\mathcal{D}^{2}B^{(2)}\epsilon^{A}_{~B}\right)
-\frac{2m\Omega^{3}}{3\sinh^{3}{\rho}}A^{(2)}\delta^{A}_{~B},
\\
\Gamma^{\eta}_{\rho\rho}&=&-1-\Omega m\frac{2+\cosh{2\rho}}\sinh{\rho}-\Omega^{2}\left(2m^{2}
(1+3\coth^{2}{\rho})+\frac{A^{(2)}}{3\sinh^{2}{\rho}}\right)
\notag \\
&&~~~~~~~~~~~~~~~~~~~~~~~~~~~~~~~~~~~~~~~~~~~~~~~~~~~
-\frac{m\Omega^{3}}{\sinh{\rho}}\left( 12m^{2}\coth^{2}{\rho}+\frac{A^{(2)}}{3}
\frac{1+5\cosh{2\rho}}{\sinh^{2}{\rho}}\right),
\\
\Gamma^{\rho}_{\rho\rho}&=&
-\Omega m\frac{\coth{\rho}\cosh{2\rho}}{\sinh{\rho}}+\Omega^{2}\frac{\coth{\rho}}{\sinh^{2}{\rho}}
\left( 2m^{2}(1-2\cosh{2\rho})+\frac{A^{(2)}}{3}\right)
\notag \\
&&~~~~~~~~~~~~~~~~~~~~~~~~~~~~~~~~~
+m\Omega^{3}\frac{\coth{\rho}}{\sinh^{3}{\rho}}\left( 2m^{2}(3\cosh{2\rho}-1)
+\frac{A^{(2)}}{3}(1+5\cosh{2\rho})\right),
\\
\Gamma^{A}_{\rho\rho}&=&-\frac{\Omega^{2}}{6\sinh^{2}{\rho}}\left(\mathcal{D}^{A}A^{(2)}
+4\coth{\rho}\epsilon^{AB}\mathcal{D}_{B}B^{(2)}\right)
\notag \\
&&~~~~~~~~~~~~~~~~~~~~~~~~~~~~~~~~~~~~~~~~~~~
-\frac{m\Omega^{3}}{6\sinh^{3}{\rho}}\left(\frac{1}{\sinh^{2}{\rho}}\mathcal{D}^{A}A^{(2)}
+8\coth{\rho}\epsilon^{AB}\mathcal{D}_{B}B^{(2)}\right),
\\
\Gamma^{\eta}_{\rho A}&=&-\frac{\Omega^{2}}{6}\left( \coth{\rho}\mathcal{D}_{A}A^{(2)}
+\frac{2\cosh{2\rho}+4}{\sinh^{2}{\rho}}\epsilon_{AB}\mathcal{D}^{B}B^{(2)}\right)
\notag \\
&&~~~~~~~~~~~~~~~~~~~~~~~~~~~~~~~~~~~~~~~~~~~~~
+\frac{m\Omega^{3}}{3}\frac{\coth{\rho}}{\sinh{\rho}}\left( 3\coth{\rho}\mathcal{D}_{A}A^{(2)}
-8\epsilon_{AB}\mathcal{D}^{B}B^{(2)}\right),
\\
\Gamma^{A}_{\rho B}&=&\coth{\rho}\delta^{A}_{~B}+\Omega^{2}\left(-\frac{5}{6}
\frac{\coth{\rho}}{\sinh^{2}{\rho}}A^{(2)}\delta^{A}_{~B}+\frac{1}{3\sinh^{2}\rho}
\mathcal{D}^{2}B^{(2)}\epsilon^{A}_{~B}\right)
+\frac{2m\Omega^{3}}{3}\frac{\coth{\rho}}{\sinh^{3}{\rho}}A^{(2)}\delta^{A}_{~B},
\\
\Gamma^{\eta}_{AB}&=&\sinh^{2}{\rho}\sigma_{AB}+2m\Omega\sinh{\rho}\sigma_{AB}
-\frac{\Omega^{2}}{6}(2A^{(2)}\sigma_{AB}-\mathcal{D}_{A}\mathcal{D}_{B}A^{(2)}
-4\coth{\rho}\mathcal{D}_{(A}\epsilon_{B)C}\mathcal{D}^{C}B^{(2)})
\notag \\
&&~~~~~~~~~~~~~~~~~~~~~~~~~
+\frac{m\Omega^{3}}{9\sinh{\rho}}(16A^{(2)}\sigma_{AB}-3\mathcal{D}_{A}\mathcal{D}_{B}A^{(2)}
+12\coth{\rho}\mathcal{D}_{(A}\epsilon_{B)C}\mathcal{D}^{C}B^{(2)}),
\\
\Gamma^{\rho}_{AB}&=&\cosh{\rho}\sinh{\rho}\sigma_{AB}+2m\Omega\cosh{\rho}\sigma_{AB}
-\frac{\Omega^{2}}{6}(2A^{(2)}\coth{\rho}\sigma_{AB}-\coth{\rho}\mathcal{D}_{A}\mathcal{D}_{B}A^{(2)}
-4\mathcal{D}_{(A}\epsilon_{B)C}\mathcal{D}^{C}B^{(2)})
\notag \\
&&~~~~~~~~~~~~~~~~~~~~~~~~
+\frac{m\Omega^{3}}{9\sinh{\rho}}(16\coth{\rho}A^{(2)}\sigma_{AB}-
3\coth{\rho}\mathcal{D}_{A}\mathcal{D}_{B}A^{(2)}
+12\mathcal{D}_{(A}\epsilon_{B)C}\mathcal{D}^{C}B^{(2)}),
\\
\Gamma^{A}_{BC}&=&{}^{(\sigma )}\Gamma^{A}_{BC}+\frac{\Omega^{2}}{\sinh^{2}{\rho}}\left(\frac{5}{6}
\delta^{A}_{(B}\mathcal{D}_{C)}A^{(2)}-\frac{1}{4}\sigma_{BC}\mathcal{D}^{A}A^{(2)}\right)
-\frac{m\Omega^{3}}{\sinh^{3}{\rho}}\left(\frac{4}{9}\delta^{A}_{(B}\mathcal{D}_{C)}A^{(2)}
+\frac{1}{9}\sigma_{AB}\mathcal{D}^{A}A^{(2)}\right),
\end{eqnarray}
%
where $\sigma_{AB}$ is the metric on unit sphere and ${}^{(\sigma)}\Gamma^{A}_{BC}$ is 
the affine connection of $\sigma_{AB}$.

\section{Definition of multipole moments}

Here we introduce the definition of multipole moments following Ref. \cite{Hansen:1974zz}. 
Let us consider the stationary spacetimes with metric $\hat{g}_{\mu\nu}$ and 
denote its timelike Killing vector as $\xi$. 
Then we can define two scalar functions $\phi_{M}$ and $\phi_{J}$ 
%
\begin{gather}
\hat{\phi}_{M} \,=\,\frac{1}{4}\lambda^{-1}(\lambda^{2}+\omega^{2}-1) \\
\hat{\phi}_{J} \,=\,\frac{1}{2}\lambda^{-1}\omega
\end{gather}
%
where $\lambda=-\xi^{\mu}\xi_{\mu}$ and $\omega$ is a twist function which satisfies 
%
\begin{eqnarray}
\hat{\nabla}_{\mu}\omega \,=\,\epsilon_{\mu\nu\rho\sigma}\xi^{\nu}
\hat{\nabla}^{\rho}\xi^{\sigma}.\label{omega}
\end{eqnarray}
%
The existence of $\omega$ is guaranteed by the vacuum Einstein equation $\hat{R}_{\mu\nu}=0$. 
The quantities with hat are associated with $\hat{g}_{ab}$. 
We define the metric $\hat{h}_{\mu\nu}$ on the hypersurface normal to $\xi$ as
%
\begin{eqnarray}
\hat{h}_{\mu\nu}\,=\,\lambda \hat{g}_{\mu\nu} +\xi_{\mu}\xi_{\nu}.
\end{eqnarray}
%
Next we consider the conformal embedding $h_{\mu\nu}=\tilde{\Omega}^{2}\hat{h}_{\mu\nu}$ 
by the function $\tilde{\Omega}$ which satisfies
%
\begin{eqnarray}
\tilde{\Omega}\tilde{=}0\,,\,
D_{\mu}\tilde{\Omega}\tilde{=}0\,,\,
D_{\mu}D_{\nu}\tilde{\Omega}\tilde{=}2h_{\mu\nu} \label{omegacond},
\end{eqnarray}
%
where $D_{\mu}$ is the covariant derivative with $h_{ab}$ and
$\tilde{=}$ means the evaluation on $\tilde{\Omega}=0$ which we introduce as spatial infinity. 
By this conformal transformation, we supposed that two scalar functions $\hat{\phi}_{M}$ and 
$\hat{\phi}_{J}$ should be transformed as $\phi_{M}=\tilde{\Omega}^{-1/2}\hat{\phi}_{M}$
and $\phi_{J}=\tilde{\Omega}^{-1/2}\hat{\phi}_{J}$. 

Then, the $2^{s}$-multipole moments are defined recursively as 
%
\begin{gather}
P^{A}\,=\,\phi_{A}\\
P^{A}_{a}\,=\,D_{a}\phi_{A} \\
P^{A}_{a_{1}\cdots a_{s}}\,=\,\mathcal{O}
\Big[
D_{a_{1}}P^{A}_{a_{2}\cdots a_{s}}-\frac{1}{2}s(s-1)\mathcal{R}_{a_{1}a_{2}}
P^{A}_{a_{3}\cdots a_{s}}
\Big]
\end{gather}
%
where $\mathcal{O}[T_{ab\cdots}]$ means the symmetric traceless part of the tensor 
$T_{ab\cdots}$ and the upper index $A=(M,J)$. $\mathcal{R}_{ab}$ is the Ricci tensor of $h_{ab}$
\footnote{
In this Appendix, we use $x^{a}$ as the coordinate on the hypersurface normal 
to $\xi$.
}. 
We call $P^{M}_{a_1 \cdots a_s}$ 
and $P^{J}_{a_1 \cdots a_s}$ the mass and spin multipole moments respectively.



\end{document}